# Energy Efficient Dissociation of Excitons to Free Charges

*Authors: Maxim Tabachnyk, Samuel L. Smith, Leah R. Weiss, Aditya Sadhanala, Alex W. Chin, Richard H. Friend, Akshay Rao*

Cavendish Laboratory, J.J. Thomson Avenue, University of Cambridge, Cambridge CB3 OHE, UK

**Introductory Paragraph**

In photovoltaics (PVs) and artificial photosynthetic systems based on excitons, the question of how energy must be lost in overcoming the Coulomb barrier, converting excitons to free charges, is fundamental, as it determines the achievable open-circuit voltage or over-potential[1-3]. Here, using transient and steady-state optical spectroscopy we study a model system, pentacene/$C_{60}$, where pentacene triplet excitons are dissociated to form charge transfer states, which we show to be degenerate in energy with the triplet excitons. We directly track these charge transfer states undergoing efficient long-range separation to free charges within 50 ps, despite a significant Coulomb barrier that we measure to be 220 meV. We model this endothermic charge separation to be driven by entropic gain. Our results demonstrate that endothermic charge separation can proceed rapidly with near unity efficiency, overcoming Coulomb barriers greater than 200meV via a gain in entropy, thus minimising loss of open-circuit voltage or over-potential. In the pentacene/$C_{60}$ system this leads here to an effective $V_{OC}$ loss of only 400 meV with respect to the energy of the triplet exciton. This suggests that excitonic photoconversion systems have no fundamental disadvantage compared to systems which generate free changes directly upon photoexcitation.

**Main Text**

In silicon and other bulk inorganic semiconductors with a high dielectric constant ($\varepsilon > 10$), photon absorption generates free charges (FCs), i.e. unbound electrons and holes, which can be extracted at the electrodes of a solar cell or used to drive redox chemistry. In contrast, in molecular systems such as biological photosynthetic complexes and organic photovoltaics (OPVs), the low dielectric constant ($\varepsilon = 3 - 4$) means that photon absorption leads to the formation of Coulombically bound excitons. These excitons must then be dissociated via charge transfer at a donor-acceptor heterojunction, to form charge-transfer states (CTS), that can then further separate to form FCs if they can surmount the Coulomb barrier which is generally considered to be 200-300 meV[1]. The question of the how much energy must be lost in overcoming the Coulomb barrier, converting excitons to FCs, is fundamental, as it ultimately controls the power conversion efficiency (PCE) achievable in photovoltaics (PVs) or the over-potential available to drive chemical processes within artificial photosynthetic systems. To frame the question differently, in the best achievable device design how much energy needs to be sacrificed to dissociate excitons into free charges? Despite decades of research, the answer to this question is ambiguous. Conventionally, as shown in **Figure 1a**, it has been thought that FCs lie significantly below the exciton in free energy and that the generation of FCs thus comes with a large energy penalty[1–3], which in OPVs, results in a loss in open circuit voltage ($V_{OC}$) of typically between 550-1000 meV with respect to the exciton energy[4]. Hence, while external quantum efficiencies (EQEs) of OPVs are close to 100 %[5,6], the large $V_{OC}$ loss limits their power conversion efficiency (PCE) to around 11 %[5].

Recently, systems have been reported in which the CTS is degenerate in energy with the photogenerated exciton, suggesting that very little loss in energy may be needed to dissociate the exciton into FCs[7,8]. But how do such 'cold' CTS, which do not have excess kinetic energy, further separate to FCs? Are FCs downhill in energy from CTS, in which case they could be accessed via charge hopping within a suitable energetic landscape – at the cost of a large $V_{OC}$ loss? Or can CTS separate while moving uphill against the Coulomb barrier? This 'ideal' case, as shown in **Figure 1a**, would result in a reduction in the loss in free energy, resulting in a reduction of $V_{OC}$ loss. And if this scenario is possible, how does this separation occur and what role do the kinetics of CTS separation and recombination play? Here, in a model system, pentacene (Pc)/$C_{60}$ bilayers, we track in real time the separation of CTS, resonantly populated from the exciton. We show that these 'cold' CTS move uphill in energy, on a 50 ps timescale, against a significant Coulomb barrier, that we measure to be 220 meV. This allow the system to achieve near unity quantum efficiency, while having an effective $V_{OC}$ loss of 400 meV, comparable to non-excitonic systems such as silicon PVs. These results shed light on the mechanism and fundamental time scale on which excitons can be endothermically be converted to free charges and suggest that excitonic photoconversion systems do not have to suffer losses in efficiency related to overcoming the Coulomb barrier.

**Figure 1b** shows the molecular structures and absorption spectra of the materials studied here. Pc undergoes efficient singlet exciton fission (SF)[9], a process by which the photogenerated spin-0 singlet exciton in converted to two spin-1 triplet excitons, each of half the singlet energy, as shown in **Figure 1c**. The rate of SF in Pc is 86 fs, outcompeting alternative decay mechanisms[9,10] such as electron transfer or singlet energy transfer. The fission-generated triplets can be efficiently dissociated at a bilayer heterojunction with $C_{60}$[10]. We choose to focus on the (Pc)/$C_{60}$ bilayer for four reasons. **(I)** This system is known to have very high quantum efficiency for charge generation, with external quantum efficiencies of 126 %[11] and internal quantum efficiencies approaching 200 %[12,13]. **(II)** As we show below, the CTS at the Pc/$C_{60}$ interface are degenerate in energy with the triplet excitons energy of 0.8-0.9 eV[14,15] and importantly, Pc/$C_{60}$ devices have been shown to have a $V_{OC}$ of 0.45 V[13,16] which corresponds to a voltage loss of only 400meV with respect to the triplet exciton energy. Thus, this system shows a remarkably low energy loss for an excitonic system. **(III)** A prerequisite to achieve low $V_{OC}$ loss is that disorder, interfaces, and defects are reduced, a condition met in the highly crystalline and rigid bilayer Pc/fullerene system[4,17]. **(IV)** As this system generates FCs from triplet excitons, it avoids non-radiative losses associated with triplet exciton formation from CTS which can reduce the $V_{OC}$ of the system[18,19]. Taken together, these points mean that this is an excellent model system to understand how much energy must be lost when converting bound excitons to free charges.

The state diagram in **Figure 1c** highlights the main states with their internal energies and processes involved in charge generation at the Pc/$C_{60}$ heterojunction after photon absorption in Pc. **(1)** Photo-excitation leads to singlet excitons ($S_1$) on Pc which rapidly undergo singlet fission into pairs of triplet excitons ($T_1$)[9]. **(2)** Triplets then undergo charge transfer at the interface with $C_{60}$, which acts as an electron acceptor, to form CTS[10], as illustrated by the e-h pair across the interface in the inset of **1c**. **(3)** These CTS, must then undergo longer range charge separation, moving away from the interface to form free charges. While the free energy cannot be increased during separation, as shown in Figure 1a, the internal energy of FC can be higher than that of CTS if the entropy of the system is increased when moving from CTS to FC (as discussed below). In devices, charges would then be extracted at suitable electrodes, but in the films studied here no extraction occurs and the charges will recombine **(4)**.

**Figure 1e** shows the energetics of the main states. The energy levels of Pc and $C_{60}$ triplet excitons were found to be around 0.85 eV and 1.1 eV, obtained by measurements of the infrared phosphorescence spectra of pristine films, in agreement with previous indirect estimations[14,15,20].

The triplet energy is slightly higher than the phosphorescence peak energy due to a molecular reorganisation energy, $\lambda$. For Pc, $\lambda \approx 35$ meV was measured by approximating the energetic difference between absorption and emission of the singlet exciton to $2\lambda$ (see SI section 3). The CT energy is estimated to 0.82 eV from absorption measurements using photo-thermal deflection spectroscopy (see SI section 2), and electroluminescence (EL) measurements on a Pc/$C_{60}$ bilayer device. The EL shows the emission from CT states formed from injected charges. These results show that the CTS populated via dissociation of the triplet exciton are in resonance with triplet excitons. This makes this system very unusual because as pointed out above, as except for a few very recently developed systems[5,6], in almost all OPV, artificial photosynthetic or biological systems[3] there is a significant energetic offset between the exciton and CTS energies. Importantly, the high quantum efficiencies[11–13] of Pc/$C_{60}$ solar cells combined with the measured energetics show that state energies and not the excitonic binding energy (> 500 meV for triplets[21]) are the relevant parameter for charge dissociation, i.e. the binding energy of the initial exciton is irrelevant to either the $V_{OC}$ or external quantum efficiency.

To investigate how it is possible to get efficient charge transfer and separation without a large loss in energy, we study the ultra-fast interfacial dynamics of Pc/$C_{60}$ using femtosecond transient optical absorption (TA) measurements. In this technique a pump pulse generates photo-excitations within the film, which induce characteristic changes in absorption. These changes are then measured at different time delays using a broadband probe pulse (see Methods for details). As shown in **Figure 1d**, we prepared multilayer (ML) samples of multiple alternating ultrathin (5 nm) layers of Pc and $C_{60}$ via thermal evaporation. The thin layer thickness, corresponding to ca. 3-4 molecular monolayers for both molecules, ensured that photo-excitations are generated very close to an interface so that the observed interfacial dynamics are not convoluted with slow diffusion of triplet excitons[10]. The multiple layers serve to amplify the optical signal of the ultrathin layers. The absorption of the ML films are shown in **Figure 1b**. In the experiments we either selectively pump $C_{60}$ at 400 nm (FWHM 10 nm, <100 fs pulse lengths) or both Pc and $C_{60}$ with a fast broadband pulse centered at 560 nm (FWHM 60 nm, <50 fs).

When charges spatially separate across a heterojunction, a local electric field (E) is created between the charges, as illustrated in the inset of **Figure 1c**. This electric field shifts the transition energies of the molecules surrounding the charges, via the quadratic Stark effect[22]. This leads to a red-shifting of the Pc absorption as shown in the inset of **Figure 1b**. Comparing the absorption with and without the electric field E ($Pc(E) - Pc$), gives a differential signal conventionally known as electroabsorption (EA), which takes the form of the derivative of the absorption spectra[23]. It has been previously shown that charge generation at the Pc/$C_{60}$ heterojunction leads to such EA signals[10]. In the experiments below, we use the EA signal to track the average spatial separation of e-h pairs.

TA spectra of the multilayer sample for both excitations are shown in **Figure 2a&b**, exhibiting a pronounced positive peak around 674 nm, characteristic of the ground state bleach (GSB) of Pc[9,10], and a broad negative signal over the full visible spectrum, a photo-induced absorption (PIA) characteristic for $C_{60}$ (see SI section 5). When pumping Pc at 560 nm, we observe a blue-shift of the Pc GSB peak position within the initial 10 ps, which is not present in measurements on pristine Pc, see inset of **Figure 2a**. The blue-shift of the peak position is a result of a superposition of the GSB with an EA signal[10]. When pumping $C_{60}$, as shown in **Figure 2b**, we initially observe a broad featureless PIA, characteristic for $C_{60}$. Within 1 ps, the Pc GSB grows in, also blue-shifting compared to the position for pristine Pc. We argue that the timescale for the growth of the pentacene GSB is the timescale for charge transfer across the heterojunction and the blue-shift of the peak again is due the growth of an EA feature.

To fully understand the interplay of the overlapping spectral features, we decompose the TA data into the spectral components of the excited species, EA and their evolution over time, using a numerical decomposition method[15,18,22] based on a genetic algorithm (see Methods). **Figure 2d** shows the TA kinetics of the ML, when pumping at 560 nm and detecting between 665-680 nm, a spectral region dominated the Pc GSB and stimulated emission (SE) from the singlet exciton. The initial drop within 100 fs has been associated with a drop in SE due to rapid fission of emissive singlet excitons into pairs of triplet excitons[9,10]. We performed the numerical decomposition after 100 fs when the majority of singlet excitons have converted into pairs of triplets.

As shown in **Figure 2c**, the TA of ML films is numerically decomposed into 3 components (solid lines). Two spectral components are assigned to excitations (excitons and charges) on Pc and $C_{60}$, respectively, due to the excellent match with spectra measured on pristine films of Pc or $C_{60}$ (dotted lines) (see SI section 5). A third spectral component matches the first derivative of the TA spectrum of pristine Pc. Since a Stark shift leads to a derivative feature in TA[22] and similar EA has been previously measured on Pc[10,24], we assign this third component to EA of Pc. Note that an increased Stark-shift leads to an increase in the amplitude and not a shift of the EA signal[22].

The kinetics associated with these features are shown in **Figure 2e.** The Pc component (black) is found to be almost flat within the first 200 ps, after which time it decays. In contrast, the $C_{60}$ component (blue) rises at early times and saturates by 10 ps. This behaviour is consistent with electron transfer from triplet excitons on Pc to $C_{60}$, leading to an increase in the total number of excited states on $C_{60}$. The total population on Pc remains constant as the electron transfer leaves behind a hole on the Pc. We note that in this spectral window, there is no substantial difference between the signal of triplet excitons or holes on Pc, as it is mainly the GSB that is being probed. The slight rise in the Pc component after 10 ps is assigned to hole transfer from excitons on $C_{60}$, leading to a small rise in the excited state population of Pc. At longer times (>200 ps) all species decay due to recombination, which is greatly accelerated in the thin layers (5 nm).

Most striking however, are the kinetics of the EA signal (red), which is caused by the electric field between e-h pairs spatially separated across the heterojunction. It is proportional to the square of the electric field between each e-h pair ($E^2$), and thus sums over the contribution of each e-h pair, regardless of their mutual orientation[22,23]. The total EA signal can increase either due to an increase in the number of charge pairs or due to an increase in the distance between the electron and hole within the pair[22]. In **Figure 2e**, it can be seen that the EA signal increases up to 50 ps, whereas the total number of excitations on both Pc and $C_{60}$ (as judged by the black and blue kinetics) do no increase significantly after 10 ps. This shows that after 10 ps the increase in EA signal is not due to an increase in the number of e-h pairs.

To confirm this, we directly measure the PIA associated with the electron on $C_{60}$, by monitoring transitions associated with the anion at 550 nm and 960 nm, **Figure 3a**. These transitions, summarised in **Figure 3b**, have been previously assigned via $C_{60}^-$ sensitisation experiments[25] and theoretical calculations[26]. The dynamics of the NIR component are numerically extracted (see SI section 7) and match that of the 550 nm component as shown in **Figure 3c**, which also shows the EA component. Consistent with the kinetics in **Figure 2e**, the charge population on $C_{60}$ stops increasing by 10 ps, whereas the EA signal increases for the first 50 ps. Thus, the increase in EA signal must be associated with the population of electrons and holes separating and moving away from the heterojunction on a 50 ps timescale.

By calibrating our time resolved measurement against steady-state measurements of the EA amplitude for Pc with a known macroscropic electric field[24], we are able to estimate the energy within the microscopic electric fields of separating charges, using the time-resolved EA amplitude shown in Figure 2d [22]. We find that 220 ±50 meV of energy is stored in the electric field for each e-h pair, as it separates over a 50 ps timescale, see **Figure 4a** and SI section 9. This is the work each e-h pair must do to overcome the Coulomb barrier.

As discussed in the SI section 8, we find that this charge separation, but not the charge transfer process, is temperature dependent. This observation of slow temperature-dependent (50 ps at room temperature) charge separation contrasts with previous reports of temperature-independent sub-100 fs charge separation. In these previously studied systems, there is a considerable energetic offset between the exciton and lowest lying CTS energies and it was proposed that this allowed ballistic motion of electrons to occur following injection into higher-lying delocalised states in fullerene aggregates[22]. This mechanism enables rapid charge separation, but requires sufficient excess (kinetic) energy for the electron to surmount the Coulomb barrier and hence comes with a $V_{OC}$ penalty. In contrast, here the Pc triplet energy is in resonance with the CTS energy, and there is no excess (kinetic) energy to drive charge separation. This means that charges are moving uphill in energy by 220 meV when separating from CTS to FCs, by gaining energy from the environment.

To understand this process, we first consider the steady-state of the system and propose a thermodynamic model[4], where the ratio of free charges to bound CTS at equilibrium is controlled by state degeneracy and the CT binding energy. As discussed in the SI section 12.1, we find that both for high excitation densities in our TA experiments ($4 \times 10^3$ lattice sites per charge pair) and for lower excitation densities under solar illumination ($10^7$ lattice sites per charge pair), the system is driven towards free charges by entropy gain, since the degeneracy of separated electron and holes is greater than that of interfacially bound CTS. Thus, if the dynamics of the system are favorable, i.e. the interplay between charge recombination and CTS separation rates, then under solar illumination full charge separation will occur against the Coulomb barrier due to entropy gain[1].

Turning to the dynamics of the system, we can model the rate of CTS dissociation with two different models. We briefly summaries the models here, while full details are provided in the SI. In the first model, we consider a simplified quantum mechanical treatment of transitions between delocalised charge eigenstates[27]. In this model when low-energy bound CTS are thermally excited above the conduction edge (CE), they undergo rapid (100 fs[22]) spatial separation to free charges. The time taken for separation is then controlled by the time needed for CTS to be thermally excited above the conduction edge. In the second model, classical Marcus dynamics of incoherent hops between localised lattice sites are modelled via a kinetic Monte-Carlo approach[28]. In **Figure 4a** the modeled dynamics are compared with measured EA associated with charge separation (full details in SI section 12.2 and 12.3) and the two models are schematically contrasted in **Figure 4b**. We use standard literature parameters for both models (static disorder 50 meV, dielectric constant 3.6, an FCC (QD model) / cubic (Marcus model) lattice constant 1.5 nm, donor acceptor distance 1.5 nm, for QD model nearest neighbor coupling $J = 25$ meV, for Marcus model $J=5-50$ meV, see SI section 12.2 and 12.3 for details and references). We find that the dynamics predicted by the quantum model provides the better match to the data, however, the Marcus-type model cannot be fully excluded. In both cases CTS can separate to FCs while moving uphill in energy against the Coulomb barrier on a 50 ps timescale.

Our results show that it is indeed possible to approach an ideal excitonic photoconversion system, converting bound excitons to FCs with minimal loss in potential, as shown in Figure 1a. Kinetics are key to achieving this, as they control whether the system can approach the local thermodynamic equilibrium. If the charge separation process is sufficiently fast and competing recombination channels for the CTS can be suppressed, then the entropic gain of free charge formation enables endothermic charge separation against a significant Coulomb barrier. In the Pc/$C_{60}$ system we directly measure this barrier to be 220 meV, which is overcome in 50 ps. The height of the barrier, controlled by the dielectric constant, is likely to be similar for all organic systems. The relatively fast time scale for charge separation is also crucial, as in many of these systems the CTS lifetimes are on the order of 1 ns[29], which could allow sufficient time for endothermic charge separation, if charges can be thermally promoted into band-like states. This implies that excitonic semiconductors, such as organics or 2D layered semiconductors[30], can function as efficiently as non-excitonic semiconductors (at room temperature).

The overall $V_{OC}$ of a PV device is controlled both by the energetics, i.e. the amount of energy lost in converting excitons to free charges and the amount of non-radiative recombination. In the Pc/$C_{60}$ system, near unity quantum efficiency is achieved while separating charges against the Coulomb barrier, i.e. the energetics of charge separation are optimised. However, almost all recombination in this system is non-radiative. Despite this, the system shows an effective $V_{OC}$ loss (with respect to the triplet energy) of only 400 meV, which is comparable to non-excitonic semiconductors. This suggests that if non-radiative recombination channels for CTS, such as recombination to lower lying triplet excitons[18,19], can be suppressed, then excitonic systems could allow for $V_{OC}$ loss approaching the best inorganic semiconductors.

## Methods

**Sample Fabrication:** Films for transient spectroscopy were fabricated on 0.13-0.17 mm thin cover glass slides. The pentacene and $C_{60}$ layers were evaporated in a vacuum (<$2\times10^{-6}$ mbar) at an evaporation rate of 0.1-0.2 A/s. The samples were encapsulated in nitrogen atmosphere (<1 ppm oxygen and water) with a second 0.13-0.17 mm thin glass slide and an epoxy glue at the edges. Films for photoluminescence (PL) measurements were deposited on and encapsulated with fused silica glass of 1 mm thickness to avoid PL from the substrate.

**Steady State Optical Measurements:** The *PL* was measured by exciting with a cw diode laser, MGL-III-532 at 532 nm. Lenses project the PL emitted to a solid angle of $0.1\pi$ onto an InGaAs detector (Andor DU490A-1.7) which has a cut-off at 1620 nm. Optical long-pass filters (for IR detection cut-off at 950 nm, for visible detection cut-off at 600 nm) before the detector ensured that there was no pump light in the PL detection and short-pass filters after the excitation laser ensured that there were no IR artefacts in the excitation line. The *electroluminescence* (EL) was measured in the same experimental configuration as the PL above, but without the excitation laser. The long-pass filter before detection had a cut-off at 950 nm. Before the start of the EL measurement, the PL from Pc (see PL setup above) around 700 nm was used to align the EL setup. The *absorption* spectra of evaporated and spin-coated films was measured with a PerkinElmer Lambda 9 UV-Vis-IR spectrophotometer. In *photo-thermal deflection spectroscopy* (PDS) a monochromatic pump beam is shone on the sample, which on absorption produces a thermal gradient near the sample surface via non-radiative relaxation induced heating. The induced refractive index gradient is enhanced by immersing the sample in an inert liquid FC-72 Fluorinert® (3M Company) with a high refractive index change with temperature. A fixed wavelength CW probe laser beam is passed through this refractive index gradient producing a deflection, which is proportional to the light absorbed in the sample at that particular wavelength, detected by a photo-diode using a lock-in amplifier. Scanning through different wavelengths gives us the complete absorption spectra with ultrahigh sensitivity.

**Transient absorption (TA) spectroscopy:** A portion of the output of a Ti:Sapphire amplifier system (Spectra-Physics Solstice) operating at 1 kHz, was used to pump a home built non-collinear optical parametric amplifier (NOPA) to generate the pump pulse at 560 nm or 580 nm (FWHM 60 nm, <50 fs). Alternatively, to generate the pump at 400 nm, a portion of the Ti:Sapphire output was frequency-doubled using a β-barium borate crystal. Another portion of the Ti:Sapphire was used to pump another home built NOPA to generate a broadband probe pulse, depending on the configuration either in the visible (520-800 nm) or near-infrared (880-1100 nm). The probe beam was split to generate a reference beam so that laser fluctuations could be normalized. The pump and probe beam were overlapped on the sample, whereas the reference did not interact with the pump. To ensure constant excitation densities over the probe region on the sample, the pump diameter on the sample was made ca. 6 times bigger than the probe diameter (80μm FWHM). The probe and reference beams were dispersed in a spectrometer (Andor, Shamrock SR-303i) and detected using a pair of 16-bit 512-pixel linear image sensors (Hamamatsu). The probe was delayed using a mechanical delay stage (Newport) and every second pump pulse was omitted using a mechanical chopper. Data acquisition at 1 kHz was enabled by a custom-built board from Stresing Entwicklunsbüro. The differential transmission (ΔT/T) was calculated after accumulating and averaging 1000 "pump on" and "pump off" shots for each data point.

**Numerical Methods:** We use numerical methods based on a genetic algorithm to decompose the overlapping spectral signatures of individual excited states and obtain their kinetics. The full details of this approach can be found elsewhere[1,2]. In summary, a large population of random spectra are generating and bred to form successive generations of offspring, using a survival of the fittest approach. The best spectra are returned as optimized solutions. For a given solution, the fitness is calculated as the inverse of the sum of squared residual with a penalty added for non-physical results. The parent spectra are selected using a tournament method with adaptive crossover. The offspring are generated using a Gaussian-function mask of random parameters. To obtain peak position of the pentacene peak around 670 nm, shown in the inset of Figure 2a, we perform a Lorentzian fit to the TA data in the surrounding spectral region (FWHM 40 nm) and associate the center of the Lorentzian with the peak position.

**Acknowledgements:** M.T. thanks the Cambridge Gates Trust for financial support. This work was supported by the EPSRC and the Winton Programme for the Physics of Sustainability. A.S. would like to acknowledge the kind support from India-UK APEX project. We thank Sam Bayliss for fabricating devices for electroluminescence measurements, and Matt Menke and Florian Schroeder for useful discussions.

**Author Contributions:** M.T. prepared samples, performed the measurements and analyzed the data. S.L.S. developed the thermodynamic and kinetic theory and was supervised by A.W.C.. L.R.W provided samples for electroluminescence measurements. A.S performed PDS measurements. R.H.F. guided the work. A.R. initiated and guided the work. This manuscript was written through contribution from all authors.

**Additional Information:** Supplementary Information is available in the online version of the paper. The data supporting this publication is available at [url].

**Author Information:** Correspondence and requests for materials should be addressed to Akshay Rao (ar525@cam.ac.uk) or Richard H. Friend (rhf10@cam.ac.uk).

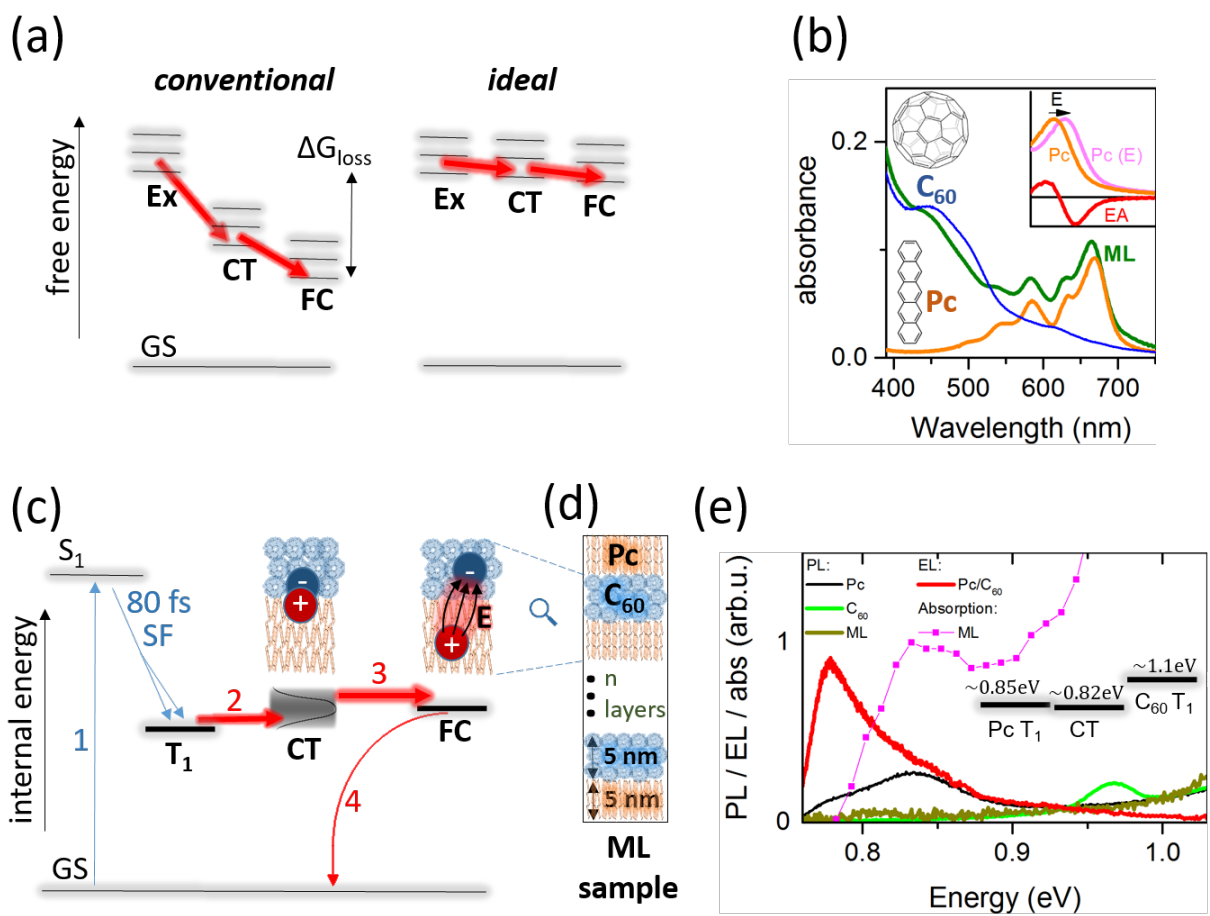

**Figure 1: Electronic processes at pentacene:fullerene heterojuntion.** **(a)** Free energy loss ($\Delta G_{loss}$) associated with the dissociation of Coulombically bound excitons (Ex) into free charges (FC) in a conventional and an ideal system. A larger free energy loss translates to a lower $V_{OC}$ in a PV device **(b)** Linear absorbance of pentacene (Pc), $C_{60}$ and multilayer (ML) films, with 15 alternating layers of 5 nm Pc and $C_{60}$. The scheme in the inset highlights how an electric field E leads to a quadratic Stark-shift in Pc absorption Pc(E). The differential signal measured is the electroabsorption (EA). **(c)** State diagram (enthalpy) showing photo-excitation (1) of singlet excitons ($S_1$) on Pc which rapidly undergo singlet fission into a pair of triplets, $T_1$. These triplets diffuse to the $C_{60}$ interface, which acts as an electron acceptor (2) to form a CTS. The charges can spatially separate (3), leading to a microscopic electric field (E). Charges can recombine (4) at the heterojunction. **(d)** To study the kinetics of interfacial processes while avoiding a convolution with slow triplet exciton diffusion, we study a ML film, comprised of alternating ultra-thin layers. **(e)** Photoluminescence (PL), absorption (abs) and electroluminescence (EL) measurements showing that the Pc triplet exciton ($T_1$) is in resonance with the CT in ML (25 alternating layers).

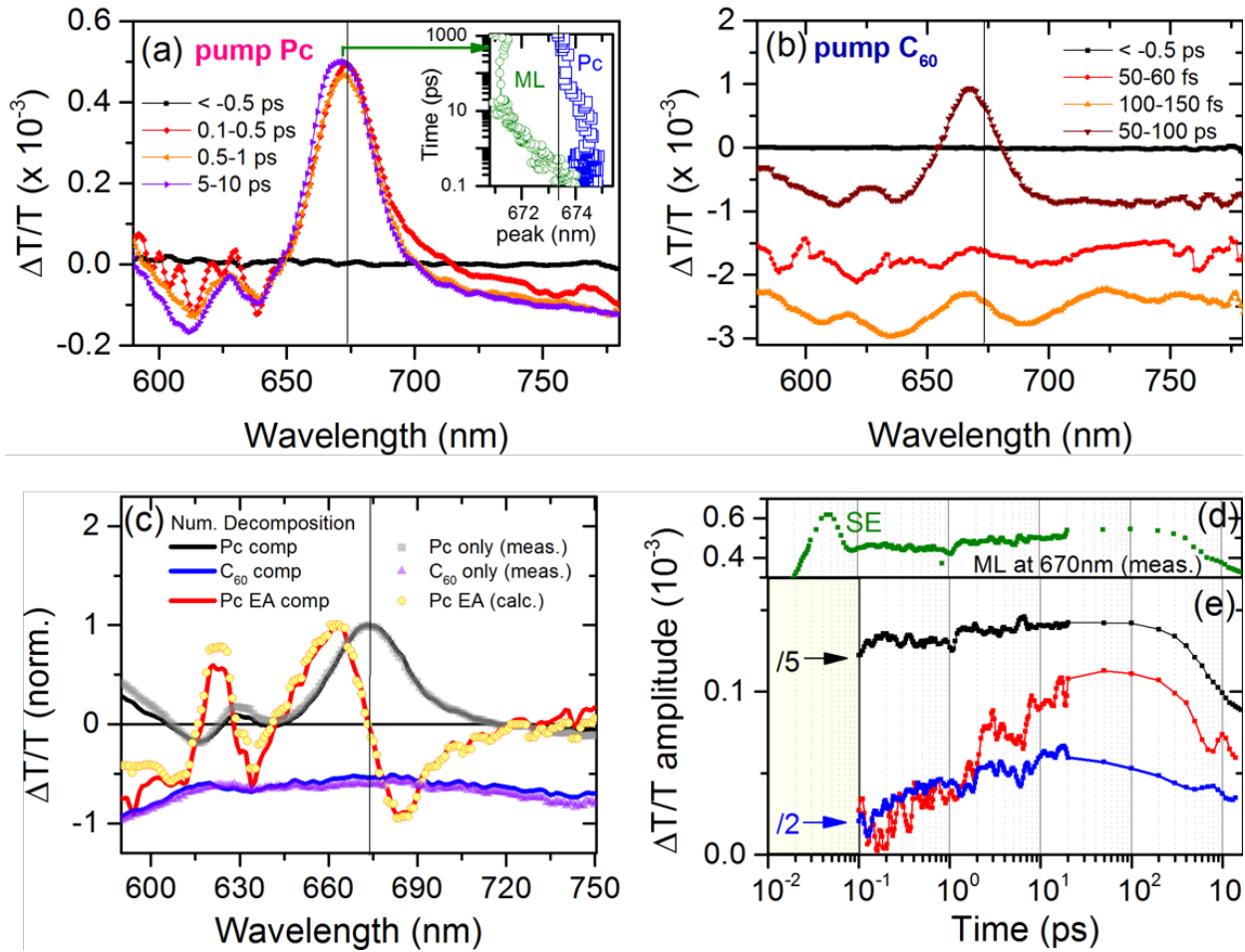

**Figure 2: Charge transfer, spatial separation and recombination dynamics.** Transient absorption **(**TA) spectra of pentacene/$C_{60}$ multilayer (ML, 15 alternating layers) films averaged over indicated pump-probe delays, **(a)** exciting pentacene (Pc) and $C_{60}$ around 560 nm or mostly **(b)** $C_{60}$ at 400 nm. The inset in (a) highlights a shift in the main ground state bleach (GSB) peak position of Pc in the first 10 ps when interfaced with $C_{60}$ (green), whereas there is no significant shift for pristine Pc (blue). The vertical line at 673.3 nm (GSB position in pristine Pc) is a guide for the eye. The TA results from (a) can be numerically decomposed into 3 components (comp). The extracted spectra and amplitude kinetics of the components are shown in solid lines in **(c)** and **(e)**. The numerically extracted spectral components are assigned to Pc and $C_{60}$ by comparison with measured spectra on pristine Pc (squares) and $C_{60}$ (triangles) films (50-300 ps & 1-2 ps). The Pc electroabsortion (EA) is calculated (circles) by differentiating the TA spectrum of pristine Pc. The sub-ps rise of the $C_{60}$ component when pumping Pc in (c) or the rise of the Pc component when pumping $C_{60}$ in (d) is associated with charge transfer. The rise of the EA component is associated with spatial charge separation. **(d)** shows the kinetics averaged over 665-680 nm for TA results presented in (a), showing an initial drop in stimulated emission (SE), indicating that most singlet excitons on Pc undergo fission within 100 fs.

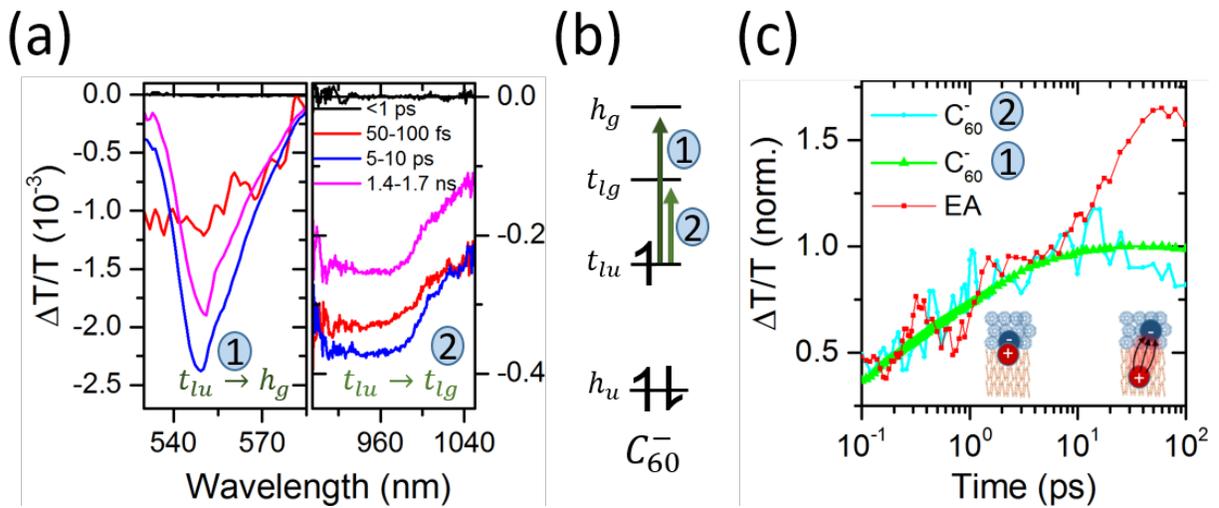

**Figure 3: Charge kinetics**. **(a)** Transient absorption (TA) spectra of pentacene/$C_{60}$ multilayer (25 alternating layers) films averaged over indicated pump-probe delays, exciting pentacene (Pc) centred at 560 nm. The photo-induced absorption (PIA) at 550 nm (1) and 960 nm (2) is associated with $C_{60}$ anion transitions indicated in **(b)**. The kinetics of the 1$^{st}$ transition and the numerically extracted (see SI) kinetics of the 2$^{nd}$ transition are compared with the electroabsorption (EA) kinetics in **(c)**. The continued rise of the EA signal beyond 10 ps indicates that spatial charge separation occurs on longer time scales than charge transfer, which as shown by the saturation of the two charge kinetics is complete by 10 ps.

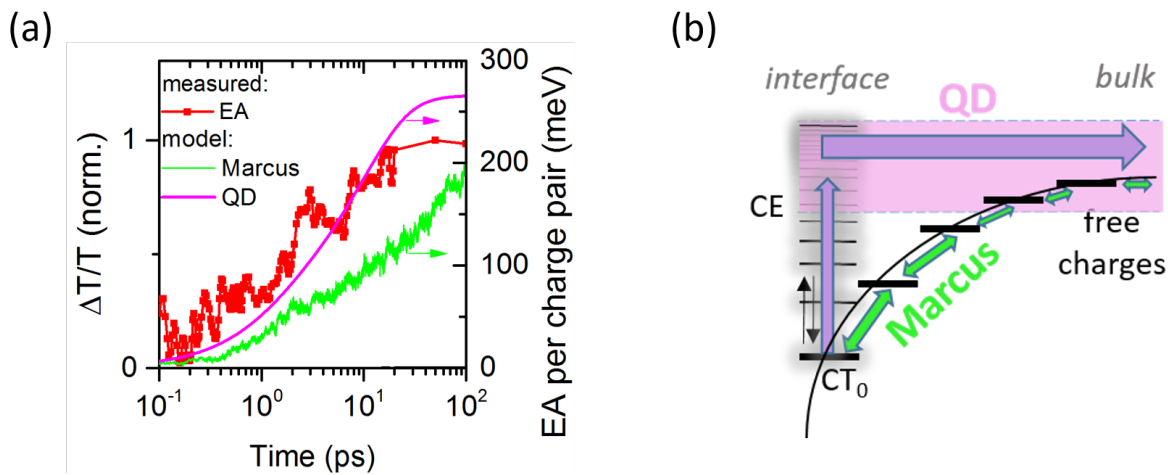

**Figure 4: Endothermic charge separation. (a)** Measured electroabsorption (EA) kinetics (red line) is correlated to the energy per charge pair and compared to modelled kinetics using either a quantum mechanical model (QD) or Marcus-type hopping, which both saturate around 100 ps. Both models are highlighted in the energetic scheme in **(b)**, showing the Coulomb barrier that charges of interfacial charge transfer states (CT) need to overcome to dissociate. In the QD model CT states are rapidly (100 fs) spatially separated when thermally excited above the conduction edge (CE), the energy required to access states delocalised beyond the Coulomb barrier, whereas the Marcus model relies on hopping of the charges up the Coulomb barrier.

# Supplementary Information

# Energy Efficient Dissociation of Excitons to Free Charges

*Authors: Maxim Tabachnyk, Samuel L. Smith, Leah R. Weiss, Aditya Sadhanala, Alex W. Chin, Richard H. Friend, Akshay Rao*

## Table of Contents

# 1. Experimental details

## 1.1 Sample Fabrication

**Film preparation for spectroscopic measurements:** Films for ultrafast transient absorption measurements were fabricated on 0.13-0.17 mm thin cover glass slides. The pentacene and $C_{60}$ layers were evaporated in a vacuum better than $2\times10^{-6}$ mbar with an evaporation rate of 0.1-0.2 A/s. The samples were encapsulated in nitrogen atmosphere (<1 ppm oxygen and water) with a second 0.13-0.17 mm thin glass slide and an epoxy glue at the edges. Films for photoluminescence measurements were deposited on and encapsulated with fused silica glass of 1 mm thickness to avoid photoluminescence from the substrate, which would otherwise mask the weak phosphorescence.

**Device fabrication for electroluminescence measurements:** Devices for electroluminescence measurements were fabricated as follows. First glass substrates coated with 150 nm ITO (used as purchased from UQG Optics) were cleaned using an ultrasonic bath of acetone followed by isopropanol and dried with nitrogen gas. Pentacene, fullerene-$C_{60}$ ($C_{60}$), and bathocuprioine (BCP) were purchased from Sigma Aldrich and then thermally evaporated in a nitrogen glovebox (<15 ppm oxygen and <1 ppm water) and stored either in nitrogen or vacuum. All materials were evaporated with a pressure below $10^{-6}$ mbar. Pentacene was evaporated at a rate of 0.4 ± 0.05 A/s with 40 nm of pentacene deposited total. Subsequent layers were deposited of $C_{60}$ (30 nm) and BCP (10 nm) at an evaporation rate of 0.1 ± 0.01A/s. Finally 100 nm aluminium electrode was deposited in vacuum (<5x$10^{-6}$ mbar) with an initial rate of 0.1± 0.1 A/s, which was increased to 0.5± 0.5 A/s after the first 10 nm. The cells were then stored and measured in a nitrogen environment (<25 ppm oxygen, <1 ppm water).

## 1.2 Transient spectroscopy

In this technique a pump pulse generates photoexcitations within the film, which are then studied at some later time using a broadband probe pulse. A portion of the output of a Ti:Sapphire amplifier system (Spectra-Physics Solstice) operating at 1 kHz, was used to pump a home built non-collinear optical parametric amplifier (NOPA) to generate the pump pulse at 560 nm or 580 nm (FWHM 60 nm, <50 fs). Alternatively, to generate the pump at 400 nm, a portion of the Ti:Sapphire output was frequency-doubled using a β-barium borate crystal. Another portion of the Ti:Sapphire was used to pump another home built NOPA to generate a broadband probe pulse, depending on the configuration either in the visible (520-800 nm) or near-infrared (880-1100 nm). The probe beam was split to generate a reference beam so that laser fluctuations could be normalized. The pump and probe beam were overlapped on the sample, whereas the reference did not interact with the pump. To ensure constant excitation densities over the probe region on the sample, the pump diameter on the sample was made ca. 6 times bigger than the probe diameter (80μm FWHM). The probe and reference beams were dispersed in a spectrometer (Andor, Shamrock SR-303i) and detected using a pair of 16-bit 512-pixel linear image sensors (Hamamatsu). The probe was delayed using a mechanical delay stage (Newport) and every second pump pulse was omitted using a mechanical chopper. Data acquisition at 1 kHz was enabled by a custom-built board from Stresing Entwicklunsbüro. The differential transmission (ΔT/T) was calculated after accumulating and averaging 1000 "pump on" and "pump off" shots for each data point.

Note that in contrast to multiple pass TA[1,2] which can be used to amplify the signal of thin layer, there is no loss in time resolution when using the multilayer approach.

## 1.3 Steady-State spectroscopy

The **photoluminescence** (PL) was measured by exciting with a cw diode laser, MGL-III-532 at 532 nm. Lenses project the PL emitted to a solid angle of $0.1\pi$ onto an InGaAs detector (Andor DU490A-1.7) which has a cut-off at 1620 nm. Optical long-pass filters (for IR detection cut-off at 950nm, for visible detection cut-off at 600nm) before the detector ensured that there was no pump light in the PL detection and short-pass filters after the excitation laser ensured that there were no IR artefacts in the excitation line. The simplicity of the setup, where the emission is directly coupled onto the detector combined with highly sensitive detectors in the infrared allow very high photoluminescence sensitivity.

The **electroluminescence** (EL) was measured in the same experimental configuration as the PL above, but without the excitation laser. The long-pass filter before detection had a cut-off at 950nm. Before the start of the EL measurement, the PL from Pc (see PL setup above) around 700nm was used to align the EL setup.

The **absorption** spectra of evaporated and spin-coated films was measured with a PerkinElmer Lambda 9 UV-Vis-IR spectrophotometer. Highly sensitive absorption measurements were further obtained with photo-thermal deflection spectroscopy as described below.

**Photo-thermal Deflection Spectroscopy** (PDS) is a highly sensitive absorption measurement technique. For the measurements, a monochromatic pump beam is shone on to the sample (film on fused silica substrate), which on absorption produces a thermal gradient near the sample surface via non-radiative relaxation induced heating. This results in a refractive index gradient in the area surrounding the front of the sample surface. This refractive index gradient is further enhanced by immersing the sample in an inert liquid FC-72 Fluorinert® (3M Company) which has a high refractive index change per unit change in temperature. A fixed wavelength CW probe laser beam is passed through this refractive index gradient producing a deflection, which is proportional to the light absorbed in the sample at that particular wavelength, which is further detected by a photo-diode and lock-in amplifier combination. Scanning through different wavelengths gives us the complete absorption spectra.

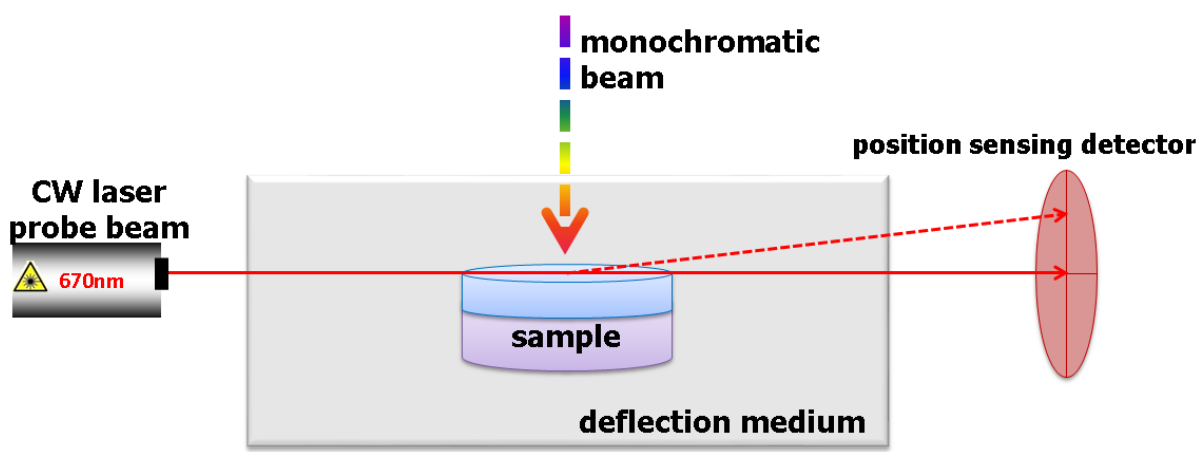

**Figure S2:** Schematic of the Photo-thermal deflection spectroscopy (PDS) setup.

## 2. Photo-thermal deflection spectroscopy

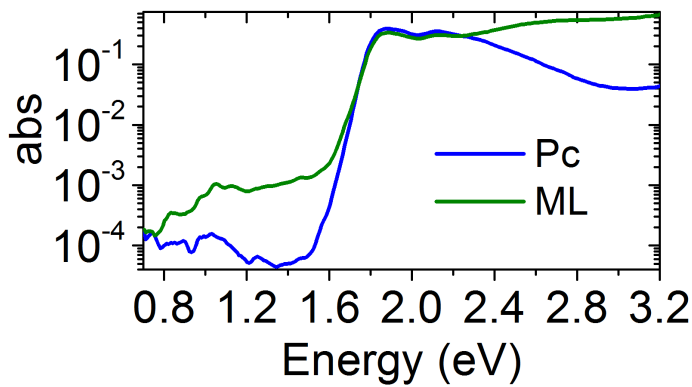

**Figure S2:** Photo-thermal deflection spectroscopy (PDS) on a pristine pentacene (Pc) film with 65 nm film thickness and a multilayer (ML) film, consisting of 13 layers of Pc and 12 layers of C60 alternating and 5 nm thickness each. The absorption spectra of Pc is found to drop sharply at the band edge, falling more than three orders of magnitude. In contrast, for the ML film there is a plateau in the drop and enhanced absorption between 0.8 -1.6eV. We assign this absorption to CTS at the Pc/$C_{60}$ interface. The measured absorption suggests that there is wide distribution of CTS energies, which is consistent with other OPV systems. There are also peaks in the absorption at 1.1ev and 0.85eV. Since the energy of the pentacene triplet state ($T_1$) is 0.85eV, it cannot be dissociated via the high lying CT states. Rather, charge transfer from Pc to $C_{60}$ must populate the low energy tail of the CT states at 0.85eV. This shows that the exciton energy and CTS energy in this system are degenerate and that the binding energy of the exciton can still be overcome, despite a negligible offset to the CTS energy.

# 3. Phosphorescence

## 3.1 Pristine $C_{60}$

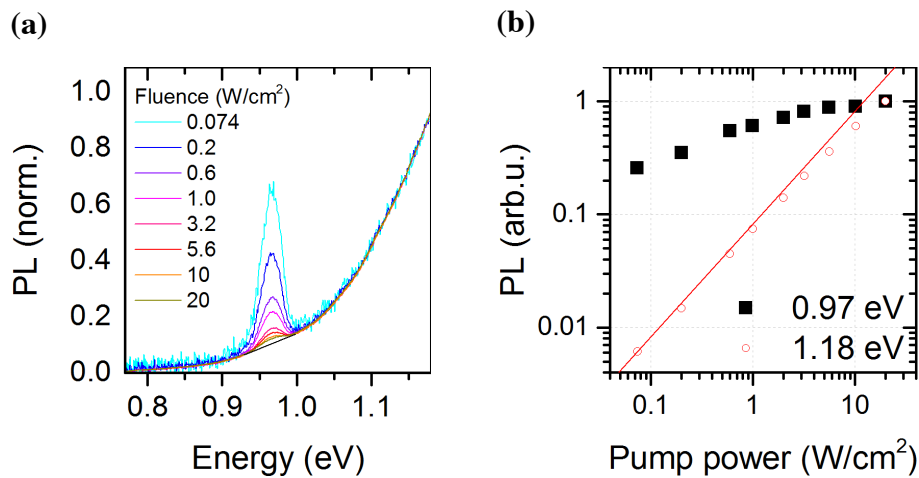

**Figure S3.1:** Photoluminescence (PL) from a pristine $C_{60}$ film of 60 nm thickness illuminated with a continuous wave laser at 532 nm. **(a)** shows the PL spectra at different fluences and **(b)** shows the normalised intensity-dependence of the singlet emission (probed at 1.18 eV) and the triplet phosphorescence (probed at 0.97 eV). The triplet phosophorescence amplitude at 0.97 eV was calculated by subtracting a polynomial fit to the singlet component, shown as a black line in (a). Whereas the singlet PL component rises linearly with pump power, matching the red line with slope 1 in (b), the triplet phosphorescence saturates at higher fluences. The latter indicates efficient triplet-triplet annihilation at higher fluences and excludes emission from immobile traps as the origin of the peak at 0.97 eV.

## 3.2 Pristine pentacene

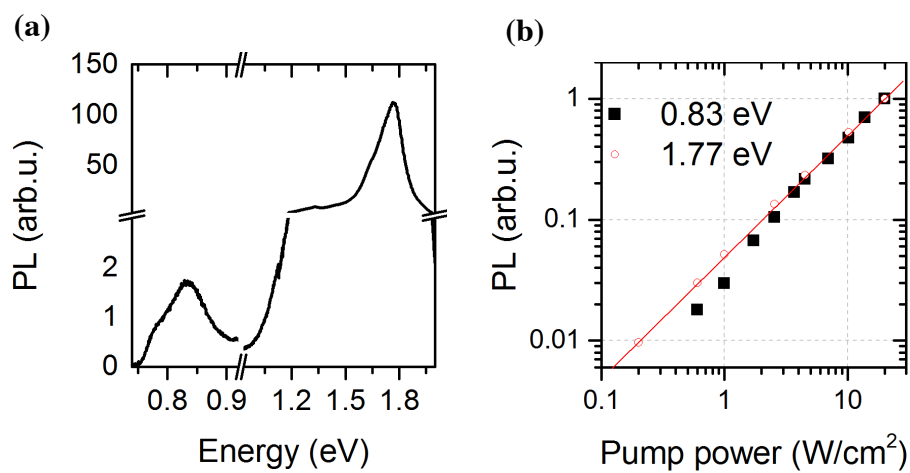

**Figure S3.2:** Photoluminescence (PL) from a pristine pentacene film of 65 nm thickness illuminated with a continuous wave laser at 532nm. **(a)** shows the PL spectrum at 20W/cm² and **(b)** shows the normalised intensity-dependence of the singlet emission (probed at 1.77 eV) and the triplet phosphorescence (probed at 0.83 eV). Both the singlet PL component and the triplet phosphorescence rise approximately linearly with pump power, matching the red line with slope 1 in (b), indicating monomolecular recombination. The latter indicates that at excitation densities in these experiments ($< 3 \cdot 10^{16} \frac{1}{cm^3}$) corresponding to an average triplet exciton distance of >25 nm, there is no efficient bimolecular triplet-triplet annihilation.

## 3.3 Reorganisation energies and triplet energy

In Figure S3.2, the PL from the Pc singlet peaks at 1.78 eV, whereas the first absorption peak is at 1.85 eV (see Fig.1a in main text), indicating a reorganisation energy $\lambda$ in Pc of around 35 meV, assuming a difference of $2\lambda$. In the main text, we assume a similar reorganisation energy for the Pc triplet to determine the triplet energy from the triplet phosphorescence. Noting that the phosphorescence energy is around $\lambda$ lower than the state energy, we estimate the triplet energy in pentacene to $0.85\pm0.03$ eV, consistent with previous calculations and measurements[3,4].

The value for the reorganisation energy of the $C_{60}$ triplet could not be determined directly. However, it is known that the reorganisation energy for charges on $C_{60}$ is as low as 50-60 meV because of the molecular size and stiffness[5]. So we assume that the reorganisation energy of the triplet has a similar magnitude and is below the reorganisation energy of two charges, 120 meV. Combined with the phosphorescence energy, we estimate the triplet energy in $C_{60}$ to 1.0-1.2 eV, consistent with previous theoretical calculations (intermediate neglect of orbital overlap model in ref. [6]) but lower than in ref. [7].

The energy of the CT state is estimated to be between the electroluminescence and absorption peak of the CT state (see Figure 1d in the main text), which is around $0.82\pm0.03$ eV.

## 4. Electroluminescence

The CT energy is estimated to 0.82 eV via ultrasensitive absorption measurements using photo-thermal deflection spectroscopy (see above). Instead of generating charge transfer states optically as in the absorption and photoluminescence measurements, we can additionally estimate the charge transfer energy by observing the emission from electrically injected charges. The result from such an electroluminescence (EL) measurement on a Pc/$C_{60}$ bilayer device is presented in the main text in Figure 1d. From the energetic difference between absorption and electroluminescence of 50-100 meV, we can estimate a low interfacial reorganisation energy of ca. 25-50meV, consistent with the low reorganisation energies found for Pc and $C_{60}$ above as well as DFT calculations for their interface in ref. [6]. Moreover, from the low width of the EL peak and the lowest-energy absorption peak of the charge transfer state, we can estimate a low interfacial energetic disorder of 20-50meV.

Importantly, we note that the EL shoulder around 0.83 eV may originate from triplet emission due to the resonance between the CT and triplet level. This suggests that non-geminate recombination of electrons and holes, in this case via electrical injection of charges, generates triplet excitons. The implication of this is that during photovoltaic operation, the recombination of charges is likely to regenerate the triplet excitons which can then be re-dissociated to generate charges. This substantially slows down recombination, allowing the system to move towards the thermodynamic equilibrium.

# 5. TA on pristine pentacene and fullerene

## 5.1 Pc only

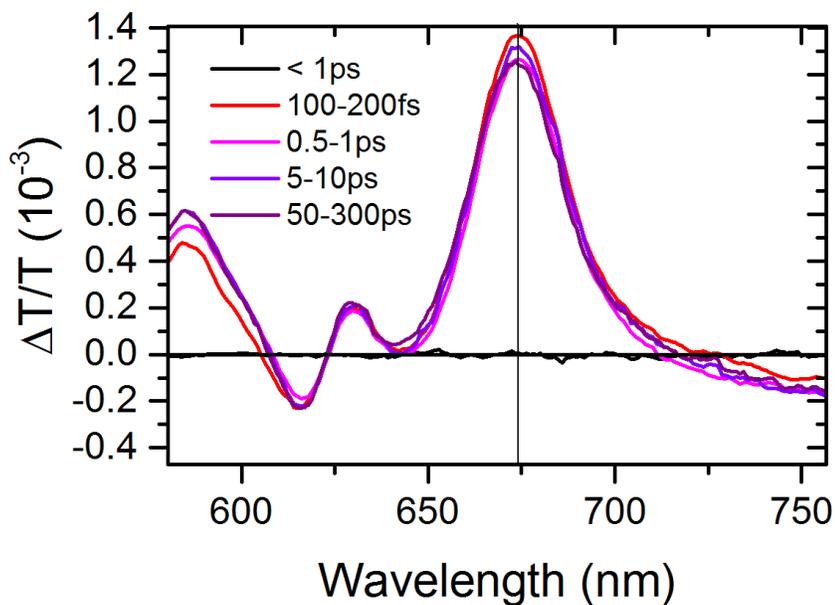

**Figure S5.1:** Transient absorption measurement on pristine pentacene film of 100 nm thickness. The excitation pulse is a fast broadband pulse around 560 nm (FWHM 60 nm, <50 fs, 14 µW/cm$^2$). The time delay between the excitation (pump) and probe pulse are indicated in the legend.

## 5.2 C$_{60}$ only

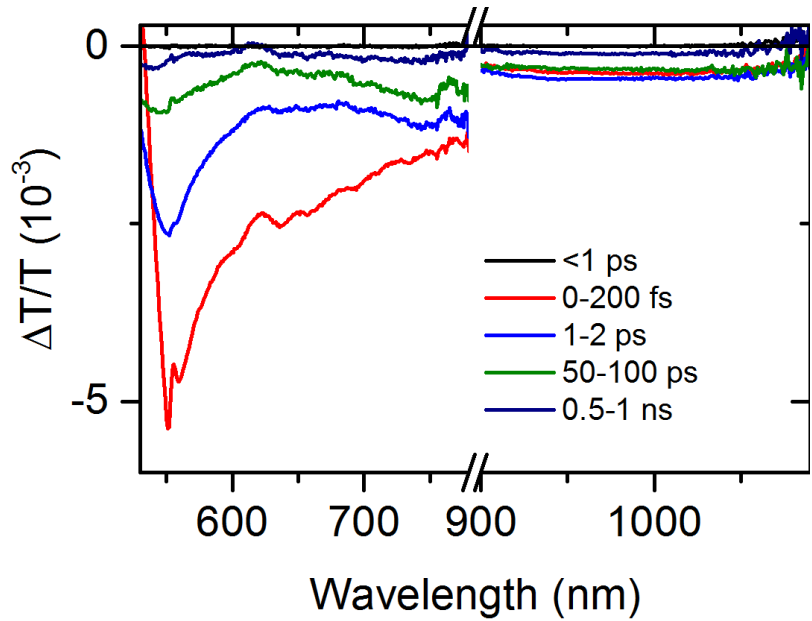

**Figure S5.2:** Transient absorption measurement on pristine C$_{60}$ film of 60 nm thickness. The excitation pulse is at 400nm (FWHM 10 nm, <100 fs, 14μW/cm$^2$). The time delay between the excitation (pump) and probe pulse are indicated in the legend.

## 6. TA on multilayer when pumping fullerene

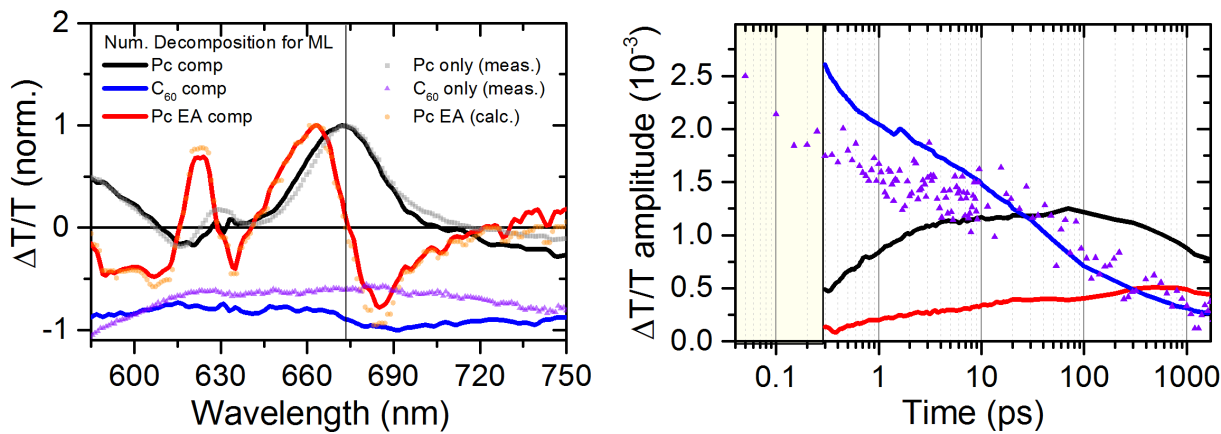

**Figure S6:** Numerical decomposition of transient absorption (TA) spectra of pentacene/$C_{60}$ multilayer (ML, 15 alternating layers) films. The measured TA spectra are presented in the main text in Figure 2b. The TA data can be numerically decomposed into 3 components (comp). The extracted spectra and amplitude kinetics of the components are shown in solid lines in **(a)** and **(b)**. The numerically extracted spectral components are assigned to Pc and $C_{60}$ by comparison with measured spectra on pristine Pc (squares) and $C_{60}$ (triangles) films (50-300ps & 1-2ps). An approximation for the Pc electroabsortion (EA) is calculated (circles) by deriving the TA spectrum of pristine Pc. The vertical line at 673.3 nm (GSB position in pristine Pc) in (a) is a guide for the eye.

## 7. TA on multilayer in NIR

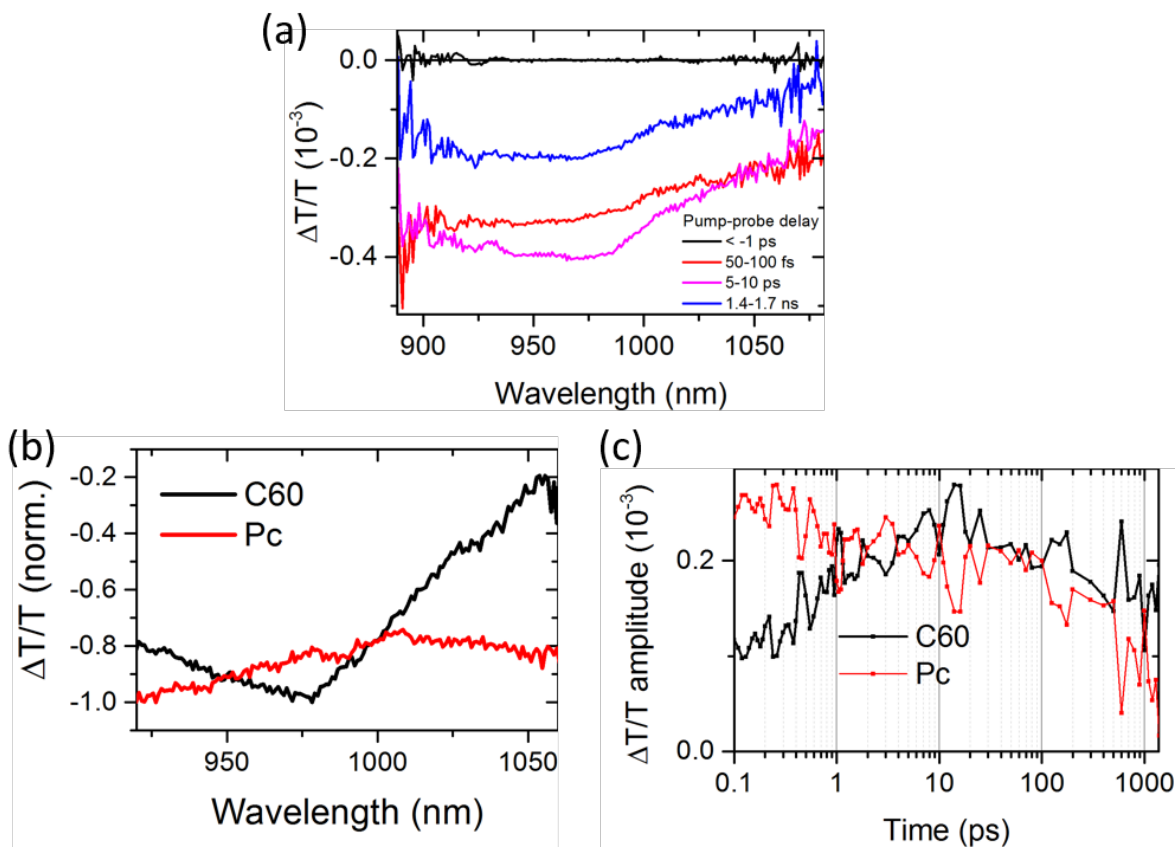

**Figure S7: Early build-up of charge PIA of $C_{60}$ in near-infrared.** (a) Transient absorption (TA) spectra of Pc/$C_{60}$ multilayer (ML, 25 alternating layers) films averaged over indicated pump-probe delays, exciting mainly Pc around 560 nm. The TA evolution can be numerically decomposed into 2 components. The extracted spectra and kinetics of the components are shown in solid lines in (b) and (c), respectively. The component peaking at 970 nm is assigned to a photo-induced absorption (PIA) of the $C_{60}$ anion ($t_{lu} \rightarrow t_{lg}$), comparing it to previous sensitisation measurements[8], theoretical calculations[9] and kinetic correlation with the charge PIA around 550 nm (see main text Figure 3a-c).

## 8. Temperature dependence of TA

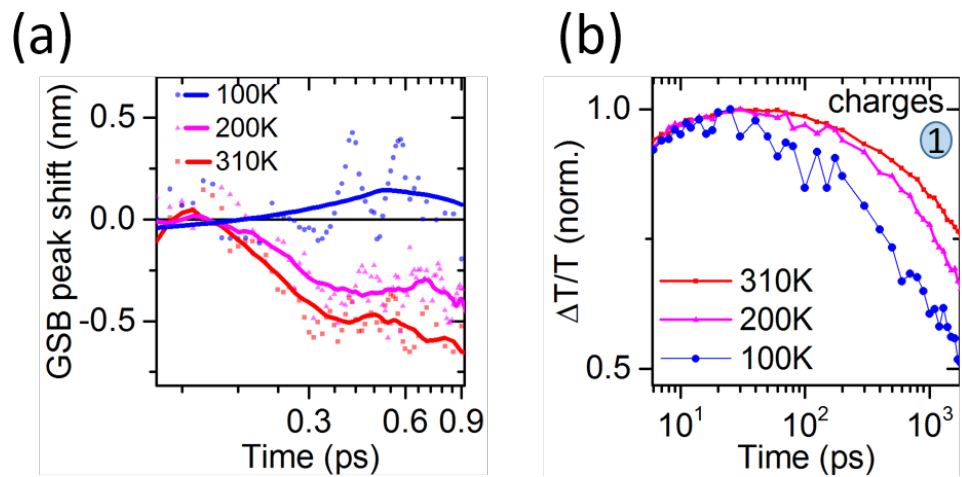

**Figure S8.1:** Transient absorption (TA) spectra of Pc/$C_{60}$ multilayer (ML, 25 alternating layers) films. **(a)** compares the shift of the Pc ground state bleach peak position around 673 nm in a ML sample, which is associated with spatial charge separation through electroabsorption. At lower temperature, the charge transfer (CT) states are less efficiently separated. The solid lines are a guide for the eye. **(b)** shows that the PIA of the $C_{60}$ anion at 550 nm (1$^{st}$ transition in Figure 3 in main text) transition decays faster at lower temperatures, indicating that charges are recombining faster at lower temperatures.

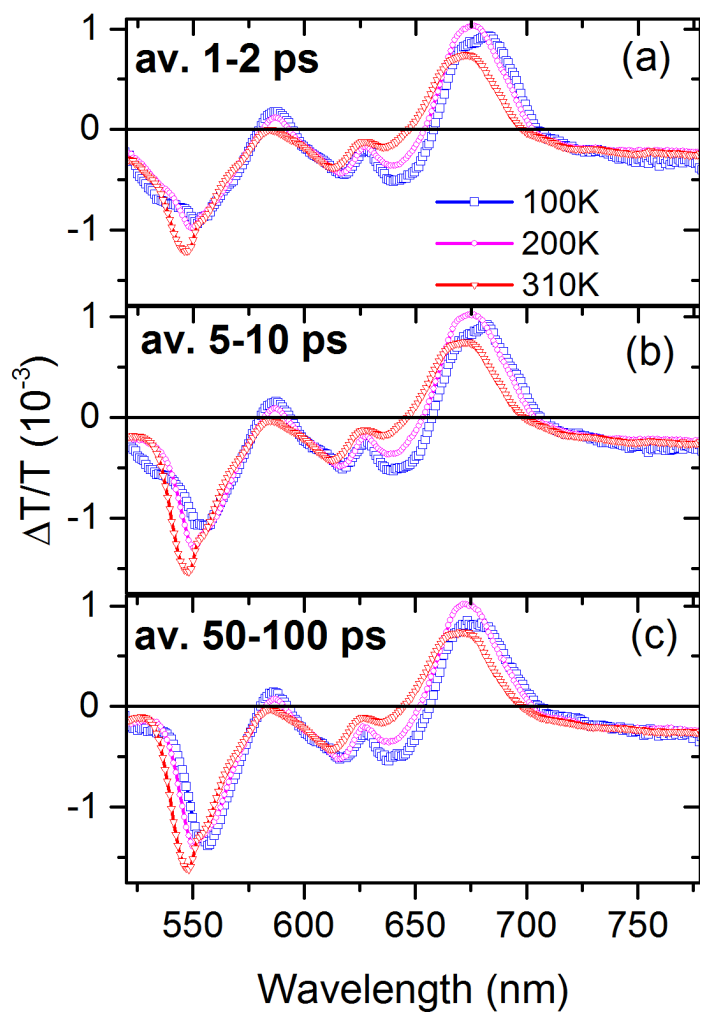

**Figure S8.2:** Transient absorption (TA) spectra of Pc/$C_{60}$ multilayer (ML, 25 alternating layers) films averaged over indicated pump-probe delays (top left in each panel), exciting mainly Pc around 560 nm. The TA is measured at 310K, 200K and 100K on the same sample and the same setup configuration. The photo-induced absorption (PIA) at 550nm, which is characteristic for the $C_{60}$ anion, grows in for all temperatures, indicating the charge transfer does not have a strong dependence on temperature.

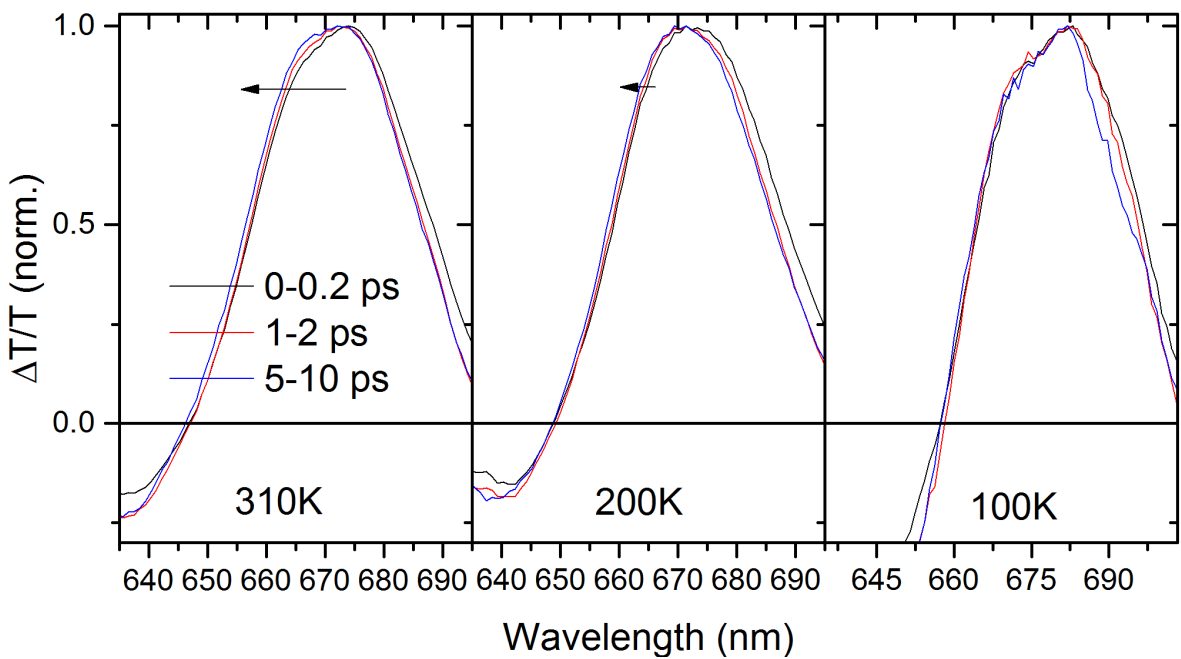

**Figure S8.3:** Transient absorption (TA) spectra of Pc/$C_{60}$ multilayer (ML, 25 alternating layers) films averaged over indicated pump-probe delays, exciting Pc around 560 nm. The TA in the three panels is measured at 310 K, 200 K and 100 K. The spectral blue-shift, which is induced by electroabsorption, increases with higher temperatures, indicating temperature-activation of spatial charge separation.

We note that the TA spectra of Pc suffers from a strong thermal artefact which becomes more pronounced at lower temperatures, as has been discussed previously[10]. This makes the numerical decomposition of the spectra at low temperatures challenging as new components needed to be added to account for the thermal artefacts. This also means that a comparison of numerically decomposed components between high and low temperatures is unreliable. Therefore, we chose to focus on the blue shifting of the Pc spectra which provides a clear measure of the electroabsorption which can be compared between different temperatures.

# 9. Calibration of macroscopic to microscopic electroabsorption

Calibration of EA amplitude at 680nm to electric field E in **macroscopic measurement**:

$$EA \propto \alpha E^2 \text{ with } \alpha = \frac{5 \cdot 10^{-4}}{50 \cdot 10^{10} \frac{V^2}{cm^2}} = 10^{-15} \frac{cm^2}{V^2} \left(\pm 0.2 \cdot 10^{-15} \frac{cm^2}{V^2}\right) \text{ (see Haas et al.}^{11} \text{ Fig. 3)}$$

In **microscopic picture** with $\varepsilon$ as permittivity and $E_{total}$ as total energy in electric field within volume V:

$$EA = \frac{1}{V} \int \alpha E^2 dE = \frac{2\alpha}{\varepsilon V} \int \frac{1}{2} \varepsilon E^2 dV = \frac{2\alpha}{\varepsilon V} E_{total}$$

Using $\varepsilon = 3.6\varepsilon_0$ (average over values in ref.[12,13]) and our measured EA amplitude of $1.15 \cdot 10^{-4}$, we find for the energy density within the microscopic electric fields in our measurement:

$$\frac{E_{total}}{V} = \frac{1}{2} \varepsilon \frac{EA}{\alpha} = \frac{1}{2} 3.6 \cdot 8.85 \cdot 10^{-12} \frac{F}{m} \frac{1.15 \cdot 10^{-4}}{10^{-19} \frac{m^2}{V^2}} = 1.83 \cdot 10^4 \frac{J}{m^3}$$

**Determination of density of charge pairs:**

Fluence $8 \frac{\mu J}{cm^2} \left(\pm 1 \frac{\mu J}{cm^2}\right)$ corresponds to a photon flux at 560nm of $2.26 \cdot 10^{13} \frac{1}{cm^2}$

12% absorption (see Figure 1a)

→ Excitation density within 75 nm: $\varrho_1 = 0.12 \cdot \frac{2.26 \cdot 10^{13} \frac{1}{cm^2}}{75 \text{ nm}} = 3.61 \cdot 10^{17} \frac{1}{cm^3}$

Correcting for fission (Pc absorption 46% of total absorption at 560nm):

$$\varrho = 1.46 \cdot 2.97 \cdot 10^{17} \frac{1}{cm^3} = 5.27 \cdot 10^{17} \frac{1}{cm^3} = 5.27 \cdot 10^{23} \frac{1}{m^3}$$

**Energy per charge pair**, assuming equal distribution of electric field energy on Pc and fullerene:

$$E_{pair} = \frac{E_{total}}{V_\varrho} = 3.5 \cdot 10^{-20} \text{J} = 217 \text{ meV } (\pm 50 \text{ meV})$$

# 10. Voltage loss

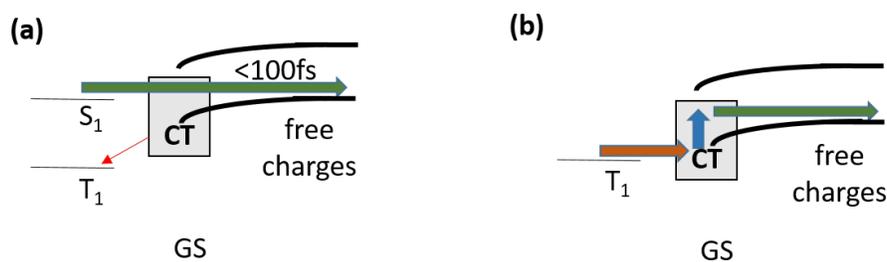

**Figure S10.1: State diagrams (enthalpy) showing how the Coulomb barrier (thick line) can be overcome when dissociating bound interfacial charges.** **(a)** In conventional high-performance OPV, a photogenerated singlet exciton ($S_1$) couples to energetically higher lying delocalised charge transfer (CT) states, which rapidly (<100 fs) lead to free charges via coherent charge delocalisation[14]. However, in this approach the excess energy of $S_1$ with respect to the energetically lowest CT state is not used (CT energy is upper bound for $V_{OC}$[15]) and lead to a lower voltage than in an optimal device where the full excitonic energy is utilised. Further, in the conventional approach charges can decay to energetically lower lying triplet excitons ($T_1$) which represents a important decay channel[16]. **(b)** In this work we investigate a system where $T_1$ is populated by photoexcitations and $T_1$ has the same energy as the lowest CT state. $T_1$ efficiently populated the CT state which efficiently separates into free charges within 100 ps. This approach avoids the potential drop due to excitonic excess energy and eliminates $T_1$ as a loss channel.

## 10.1 Charge transfer

In order to provide a driving force for charge transfer, the electron affinities of the donor and acceptor in polymer/fullerene blends are typically chosen to have a difference of 100-300 meV (see ref.[17]). This leads to a surplus of excitonic energy compared to the energetically relaxed interfacial charge transfer states, which was found to allow ultrafast spatial charge separation after charge transfer via charge delocalisation mediated by energetically excited charge transfer states[14].

As described in the main text, the pentacene triplet is in resonance with the interfacial CT state, resulting in a voltage gain of 100-300 mV voltage gain compared to typical OPV.

### 10.2 Charge separation and recombination

Burke et al.[17] derived the following expression for the V$_{OC}$ of OPVs, assuming a thermodynamic equilibrium between the interfacial charge transfer state (CT) and the free charges, as well as Gaussian density of states distribution of CT.

$$qV_{OC} = E_{CT} - \frac{\sigma_{CT}^2}{2kT} - kT \ln\left(\frac{qfN_0L}{\tau_{CT}J_{SC}}\right) \quad (1)$$

with q being the elementary charge, $E_{CT}$ the energy of CT, $\sigma_{CT}$ the energetic CT state distribution, k the Boltzmann constant, T the temperature (assumed as room temperature in the following), f the interface/volume ratio in the active layer, $N_0$ the density of electronic states in the device, L the active layer thickness, $\tau_{CT}$ the lifetime of CT states and $J_{SC}$ the short-circuit current.

First, we consider the second term in eq. (1) containing the energetic state distribution $\sigma_{CT}$. For the observed $\sigma_{CT} = 25 - 50$ meV (see SI section 4), we find $\frac{\sigma_{CT}^2}{2kT} = 13 - 50$ meV for a pentacene/C$_{60}$ interface. However, this loss is 75-225 mV for polymer/fullerene bulk heterojunction devices[17]. So the voltage loss through interfacial disorder is ca. 25-210 mV smaller in Pc/C$_{60}$ devices compared to polymer/fullerene blends.

We now turn to a comparison of the term $kT \ln\left(\frac{qfN_0L}{\tau_{CT}J_{SC}}\right)$ from equation (1) in typical Pc/C$_{60}$ bilayer solar cells and polymer/fullerene blends. The main difference between Pc/C$_{60}$ bilayer devices as in ref. [18–20] and polymer/fullerene blends is f, being approximately 10-25 times smaller for these bilayers. The lower f results in 60-80 mV less voltage loss Pc/C$_{60}$ bilayer devices compared to highly mixed blend devices.

Note that due to carrier multiplication in singlet fission, pentacene is capable of doubling the photocurrent ($J_{SC}$). For highly efficient devices with the current predominantly generated from pentacene, an additional voltage gain of $kT\ ln(2) = 20$ mV could be achieved. However, in real pentacene devices in ref. [18–20], >100% quantum efficiencies could only be achieved in a small spectral region. Therefore, $J_{SC}$ resulting from the full sun spectrum is not doubled in these devices so that singlet fission does not increase the voltage significantly in these devices.

A significant loss channel for CT states in polymer/fullerene blends is the recombination to triplet states[16,21]. Here this problem is overcome by generating charges directly via triplets. But it has been previously shown that under optimal conditions recombination to triplets can be shut down[22].

So the low interfacial disorder and bilayer design give Pc/$C_{60}$ bilayer solar cells a voltage enhancement of around 85-290 mV, compared to polymer/fullerene blends, and further enhancement is achieved by blocking the triplet recombination channel. In summary, the low voltage loss in pentacene/fullerene solar cells is consistent with theoretical considerations in ref. [17].

The estimation in the main text of the potential to obtain 20 % PCE in OPVs relies on the assumption that all solar cell characteristics of the currently best OPV[23] devices can be retained while reducing the voltage loss to 0.4 V.

### 10.3 General discussion

A prerequisite to achieve the low $V_{OC}$ loss is that disorder, interfaces, and defects[15] are reduced in the highly crystalline and rigid bilayer Pc/fullerene system, supported by our results for low interfacial disorder (FWHM of EL in Figure 1d in the main text) and previous calculations[6].

Combining the discussed effects in charge transfer, separation and recombination results in a voltage gain of 185-590 mV of Pc/C$_{60}$ bilayer solar cells compared to conventional OPV.

We note that the low photoluminescence quantum yield ($PLQE$) of pentacene singlet excitons ($PLQE < 10^{-4}$) results in a potential drop[24] of the singlet of $k_B T ln(PLQE) \approx 230$ mV. Since singlet excitons split into a pair of triplets, this results in a potential drop of ca. 115 mV for triplet excitons.

# 11. Novel photovoltaics design rules

From our work, some possible photovoltaics design rules arise, allowing to efficiently harvest excitonic energy.

Whereas triplets are generated via singlet exciton fission in our model system using pentacene, they could be generated via alternative mechanisms such as intersystem crossing in systems with small singlet-triplet energy gaps, such as those currently used in thermally activated delayed fluorescence (TADF) systems[25]. TADF systems are know to have long lifetimes associated with the CT character of the photoexcitation. This may lend itself well to generating long lived CTS with a suitable acceptor. Thus the system could allow effective entropically driven charge separation via the use of long-lived CTS.

TADF systems would further have the advantage of a strong coupling of interfacial charge transfer states (CTS) to luminescent singlet exciton states. This coupling potentially makes the majority of charge recombination to be radiative which could lead to a further increase in voltage, as observed for going from silicon to gallium arsenide in solar cells[24]. Compared to spin-singlet excitons, triplets often have the advantage of longer lifetimes and diffusion lengths, allowing for longer range energy transport, e.g. from the point of generation to the point of dissociation. Another advantage is that loss through recombination from the charge transfer state to the triplet state, which is typically occurring in OPV[16,21], can be avoided by efficiently harvesting triplets.

The second option is to harvest bright singlet excitons directly. Here, the system must be designed to ensure resonance between the singlet exciton and the energetically relaxed charge transfer state, as has been achieved recently in some systems[26,27]. In this design, loss from the charge transfer state to triplet excitons needs to be avoided. One option is to achieve this, is to also efficiently re-separate triplet excitons into charges. Due to the energetics of the system, the exchange energy needs to be very small, so that triplets are isoenergetic with singlets and the charge transfer state is energetically accessible from the triplet. Another option to avoid loss via triplets, is to reduce the transfer rate from charge transfer state to triplet states by molecular design at the interface[16]. Another crucial requirement for the success of this approach is to ensure that the states within the singlet manifold have a lifetime which is long enough so that the charge separation mechanism discussed in this work is faster than CTS recombination.

Both above approaches rely on efficient spatial charge separation of energetically relaxed interfacial charge transfer states. As described in the main text and above, this can be achieved on sub-ns timescale in OPVs.

# 12. Theory

## 12.1 Free charge population at equilibrium

The free energy of $n$ bound CT states at equilibrium, $F_{CT}(n) = U_{CT} - TS_{CT}(n)$. Similarly the free energy of $m$ free charge pairs, $F_{free}(m) = U_{free} - TS_{free}(m)$. The internal energies $U_{CT}(n) = nE_{CT}$ and $U_{free}(m) = m(E_{CT} + E_b)$, where $E_{CT}$ is the CT state energy and $E_b$ is the binding energy. The entropy $S_{CT}(n) = k_B \ln(\Omega_{CT}(n))$, where $\Omega_{CT}(n)$ denotes the number of ways we can place n bound CT states in the system. A similar expression applies for the entropy of free charges.

To proceed, we assume that the blend is composed of alternating layers of donor and acceptor semiconductor. Both donor and acceptor have primitive cubic unit cells with the same lattice constant, and each layer is d unit cells thick, as sketched in **Fig. S12.1a**. We assume that the device contains N charge pairs in total, which may be bound or unbound, and that the total number of lattice sites in the device $V = N\varrho^{-1}$. By construction, $\varrho$ is the density of charge pairs, measured in units of the lattice constant.

We assume that bound CT states can only form between neighbouring sites at an interface between donor and acceptor semiconductor. There are $N\varrho^{-1}/d$ possible CT states, and so the total number of configurations, $\Omega_{CT}(n) = \left(\frac{N\varrho^{-1}}{d}\right)^n / n!$. We must divide by $n!$ since the electron-hole pairs are indistinguishable. A free electron or hole can occupy $N\varrho^{-1}/2$ sites, since electrons can only occupy acceptor sites and holes can only occupy donor sites. The number of free charge configurations, $\Omega_{free}(m) = \left(\frac{N\varrho^{-1}}{2}\right)^{2m} / (m!\, m!)$. Here we divide by $m!$ twice, since both electrons and holes are indistinguishable.

The chemical potential of bound charges, $\mu_{CT}(n) = \partial F_{CT}(n)/\partial n$, and similarly for free charges. Applying Stirling's formula, $\ln(n!) \approx n\ln(n) - n$, we obtain

$$\mu_{CT}(n) = E_{CT} - k_B T \ln\left(\frac{N\varrho^{-1}}{dn}\right),$$

$$\mu_{free}(m) = E_{CT} + E_b - 2k_B T \ln\left(\frac{N\varrho^{-1}}{2m}\right).$$

We define f to be the fraction of unbound charge pairs. By definition, m = Nf and n = N(1-f). Thus,

$$\mu_{CT}(f) = E_{CT} - k_B T ln\left(\frac{\varrho^{-1}}{d(1-f)}\right),$$

$$\mu_{free}(f) = E_{CT} + E_b - 2k_B T ln\left(\frac{\varrho^{-1}}{2f}\right).$$

At equilibrium, $\mu_{CT}(f) = \mu_{free}(f)$. This implies

$$f^2 + xf - x = 0, \text{ where } x = (\varrho^{-1}d/4)\, e^{\frac{-E_b}{k_B T}} \quad (1)$$

Since $f$ must lie between 0 and 1, the only valid solution is $f(x) = \frac{x}{2}\left(\sqrt{1+4/x} - 1\right)$. We plot this equation, as a function of the effective parameter x, in **Fig. S13.1b**. We see that a transition from bound CT states to free charges occurs as x increases. Note that $f(1/2) = 1/2$.

We note that equation (1) is equivalent to $\frac{f^2}{1-f} = x$. So for $f \ll 1$, $f(x) = \sqrt{x} = (\varrho^{-1}d/4)\, e^{\frac{-E_b/2}{k_B T}}$, implying that the effective activation energy necessary is half the binding energy $E_b$.

Note that we have made two significant assumptions above. First, we have assumed that the charge density is low. This enabled us to calculate the number of possible configurations of free charges and CT states, without explicitly imposing the restriction that a charge cannot occupy a site already occupied by another charge. This approximation will be valid if $\varrho^{-1} \gg 1$. In Figure 4a of the main text, we plot $f(x)$ as a function of $E_b$ at 300 K for our multilayer sample with $d \sim 4$. In our TA experiments we estimate $\rho^{-1} \sim 4\times 10^3$ lattice sites per charge pair, whereas for solar illumination we estimate $\rho^{-1} \sim 10^7$. The result for the expected fraction of free charges in our experiments are plotted in Figure S12.1 C.

Second, we have neglected the long-ranged nature of the Coulomb interaction between electron and hole. In reality, many free charge states still experience a Coulomb potential and are partially bound. Introducing this interaction would decrease the number of CT states at equilibrium, but concurrently introduce a number of partially bound "free charges", separated by a few lattice sites across the interface. Introducing this term is theoretically challenging, and a topic for future research. However it will not affect the key physics of the system, whereby a transition from bound to free charges is observed as $x = (\varrho^{-1}d/4)\, e^{\frac{-E_b}{k_B T}}$ increases.

To introduce disorder, it is convenient to use a slightly modified derivation. We recall that the probability a state is occupied at equilibrium, $P(E) = e^{-\frac{(E-\mu)}{k_B T}}$. As shown above, there are $N\varrho^{-1}/d$ possible CT states in the system. If the CT state exhibits a Gaussian disorder distribution, then

$$n = N(1-f) = \frac{N\varrho^{-1}}{d} \cdot \frac{1}{\sqrt{2\pi\sigma_{CT}^2}} \cdot e^{-\frac{(E_{CT}-\mu)}{k_B T}} \int_{-\infty}^{\infty} e^{-\frac{x^2}{2\sigma_{CT}^2}} \cdot e^{-\frac{x}{k_B T}} dx = \frac{N\varrho^{-1}}{d} \cdot e^{-\frac{(E_{CT}-\mu)}{k_B T}} \cdot e^{\frac{\sigma_{CT}^2}{2k_B^2 T^2}}$$

Rearranging, $\mu = E_{CT} - \frac{\sigma_{CT}^2}{2k_B T} - k_B T \ln\left(\frac{\varrho^{-1}}{d(1-f)}\right)$. Meanwhile the free charge state is composed of two particles, the electron and hole, each of which can occupy one of $N\varrho^{-1}/2$ states. Thus,

$$m^2 = (Nf)^2 = \frac{N\varrho^{-1}}{2} \cdot e^{-\frac{(E_{free}-\mu)}{k_B T}} \cdot e^{\frac{\sigma_e^2}{2k_B^2 T^2}} \cdot e^{\frac{\sigma_h^2}{2k_B^2 T^2}}$$

in which we have exploited the relations $E_{free} = E_e + E_h$ and $\mu = \mu_e + \mu_h$. Recalling that $E_{free} = E_{CT} + E_b$, we obtain $\mu = E_{CT} + E_b - \frac{\sigma_e^2 + \sigma_h^2}{2k_B T} - 2k_B T \ln\left(\frac{\varrho^{-1}}{2f}\right)$. Solving for $f$ we again obtain $f(x) = \frac{x}{2}\left(\sqrt{1 + 4/x} - 1\right)$, where now $x = (\varrho^{-1}d/4) e^{\frac{-E_b}{k_B T}} \cdot e^{\frac{\sigma_e^2 + \sigma_h^2 - \sigma_{CT}^2}{2k_B^2 T^2}}$. The effect of disorder is controlled by the quantity $A = \sigma_e^2 + \sigma_h^2 - \sigma_{CT}^2$. If the disorder experienced by free charges exceeds the disorder of CT states, then $A > 0$. In this case disorder drives equilibrium towards free charges. If the disorder experienced by CT states exceeds the disorder of free charges, then $A < 0$, and equilibrium shifts towards the bound CT states. A-priori, we expect the disorder of free charges and CT states to be similar, such that $A \approx 0$. For this reason we neglected disorder in the main text.

The dominant source of charge decay in the bulk heterojunction is CT state recombination. In Figure S12.1 D, we modelled the overall rate of charge decay by assuming,

$$\frac{d\rho}{dt} = \rho \cdot f_B \cdot R_{CT} = \rho \cdot (1 - f(x)) \cdot R_{CT}$$

where $f_B$ denotes the fraction of charges bound within a CT state, and $R_{CT}$ denotes the rate at which bound CT states recombine, which we assume is temperature independent. A temperature dependence arises from $f(x) = f\left(\frac{\varrho^{-1}d}{4} e^{\frac{-E_b}{k_B T}}\right)$. The temperature dependent charge recombination times (black squares in Figure S12.1 D) can be fit (solid black line) using $R_{CT} = 0.3$ ns$^{-1}$ and $E_b = 215$ meV. This value of $R_{CT}$ is significantly slower than the experimentally measured rate of CTS dissociation, red line in **Figure 4c**, allowing the system to move towards equilibrium and where free charges dominate.

Intuitively, as the temperature rises, the fraction of charges bound within CT states falls, and consequently the overall rate of recombination is reduced. This simplified model of charge recombination via the CT state neglects alternative charge recombination mechanism such as charge exciton annihilation, which has an effect on the charge density due to the charge-exciton equilibrium and has previously been observed for nanostructured pentacene/fullerene blends[28]. Exciton-charge recombination might be enhanced in our thin multilayer samples due to strong spatial confinement. Further the model assumes temperature-independence of the CT ecombination rate, which is also questionable since at higher temperatures free charges can re-form spin-singlet CT states, which presumably have a faster decay rate to the ground state than spin-protected triplet CT states that do not dissociate at low temperatures.

Our experimental data shows that the recombination timescale increases by a factor of 2.3 going from 100 K to 300 K. Since we saw no EA at 100 K, indicating that the system is dominated by CT states, this suggests that more than half of the charge pairs in our device are unbound at 300 K in our TA experiment.. If we operated the device at 300K under one sun($\rho \sim 10^{-7}$), we would expect $x \sim \frac{10^7}{2 \times 4000}$. This implies that roughly 0.08 % of charges would be bound in equilibrium at one sun. Given that we estimate a CT recombination timescale $\tau_{CT} = 1/R_{CT} \sim 3$ ns, this would imply an overall recombination timescale of approximately 4 μs in device operation, if equilibrium is reached.

We note that, since x depends on the charge density, the equation above does not predict a pure exponential decay. However the linear dependence of x on the inverse charge density is significantly weaker than its exponential dependence on inverse temperature. Since we extract the rate of recombination from the decay of ρ over a narrow range, it is reasonable to approximate the decay with an exponential.

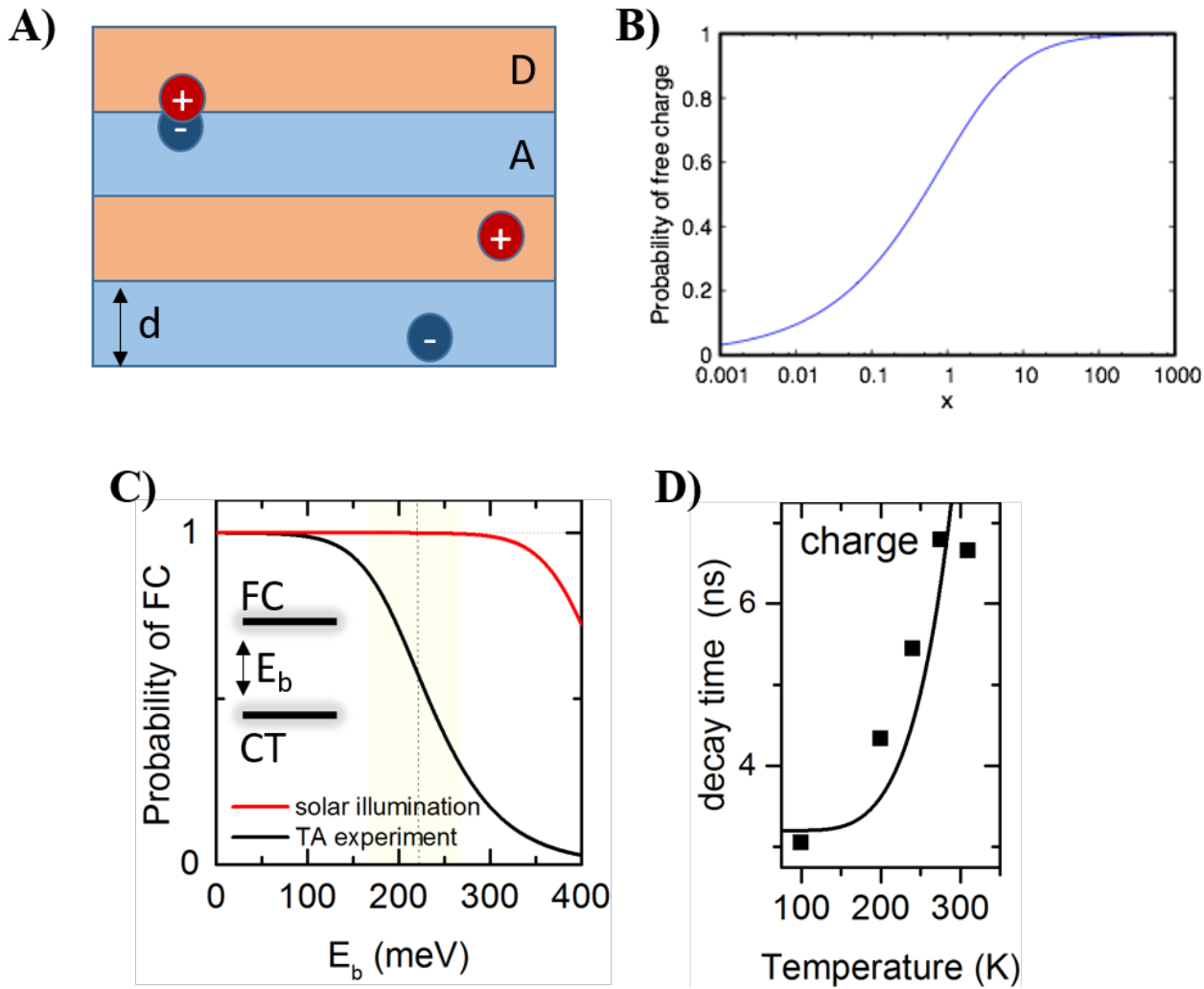

Figure S12.1: A) Our blend comprises an alternating multilayer of donor and acceptor layers. Each layer is d unit cells thick. B) The probability a charge pair is unbound is controlled by the effective parameter $x = (\varrho^{-1}d/4)\, e^{\frac{-E_b}{k_B T}} \cdot e^{\frac{\sigma_e^2 + \sigma_h^2 - \sigma_{CT}^2}{2 k_B^2 T^2}}$. When x < 1/2, there are more bound CT states than free charge pairs. When x > 1/2, there are more free charge pairs than bound CT states. C) Probability for free charges (FC) in the pentacene/$C_{60}$ multilayer in thermodynamic equilibrium with bound charge transfer states (CT). Both for low excitation densities at solar illumination and at high excitation densities as in our transient absorption measurements, the majority of charges is free for the estimated binding energy $E_B$ of 220meV (dashed line). D) Decay time of charges in pentacene/$C_{60}$ multilayer, extracted with a mono-exponential decay fit (squares) to data presented in Fig. 3e after 200ps. The temperature dependence can be modeled (solid line), assuming charges recombine via the bound interfacial state. Both for high excitation densities in our TA experiments

($4\times10^3$ lattice sites per charge pair) and for lower excitation densities under solar illumination ($10^7$ lattice sites per charge pair), the system is driven towards free charges by entropy gain.

## 12.2 Can the system reach equilibrium?

In order for the system to reach equilibrium between CT states and free charges, the rate at which CT states dissociate into free charges must significantly exceed the rate at which CT states recombine[17,26]. We found experimentally that CT states recombine on a timescale of a few nanoseconds, while the EA signature, which tracks the dissociation of CT states into free charge pairs, rises over a period of 10-100 ps. This observation justifies the claim that the charge pairs reach equilibrium between bound and unbound charges. To investigate whether such a timescale is plausible, in the following two sections we study the charge separation of bound CT states within both a quantum and a classical dynamical model. In both cases we observe the dissociation of CT states into free charges within a period of 10-100 ps, consistent with our experimental results. This further supports our experimental findings. Note that we are not seeking to compute exact dynamics, but simply to confirm that CT state dissociation within 10-100 ps is plausible, and can be obtained without assuming improbable material parameters.

## 12.3 CT state dissociation within delocalised quantum dynamics

### 12.3.1 A simple tight binding model of fullerene aggregates

In a previous work, we introduced a simple tight binding model to describe the delocalised electronic states of fullerene aggregates at the interface immediately after the exciton dissociates[14]. We neglected vibronic processes on femtosecond timescales, and assumed that the hole lies in a localised state on a donor molecule neighbouring the aggregate, while the electronic states can delocalise. The electronic Hamiltonian $H_S = \sum_i E_i |i\rangle\langle i| - J \sum_{ij}^{nn} |i\rangle\langle j|$, where the site energies $E_i = \sigma_i - q^2/4\pi\varepsilon_0\varepsilon_r r_i$. $\sigma_i$ describes Gaussian distributed static disorder, q labels the electron charge, and $r_i$ labels the distance between the $i^{th}$ lattice site and the hole. The second summation is only performed over nearest neighbours within an FCC lattice, and J labels the transfer integral. The hole lies a distance $r_{CT}$ from its nearest neighbour at the centre of one face of the aggregate. We showed that the states of this aggregate could be described by a simple band diagram, illustrated in **Fig. S12.2a**. The energies of electronic eigenstates fall near the hole at the interface, but if the bandwidth B = 16J exceeds the depth of the Coulomb well W = $q^2/4\pi\varepsilon_0\varepsilon_r r_{CT}$, then a set of fully delocalised states survive which can drive ultrafast charge separation on femtosecond timescales. These states lie above the conduction edge CE, which lies at the energy $E_{CE}$ = -12J.

On long timescales this model breaks down, and we must consider vibronic relaxation. In another work[29], we showed that trapped CT states which form on long timescales can be described by an effective Hamiltonian, $H_{eff} = \sum_i (E_i - \Delta|C_0^i|^2)|i\rangle\langle i| - J\sum_{ij}^{nn}|i\rangle\langle j|$. $\Delta$ labels the reorganisation energy, and $|C_0^i|^2$ labels the probability density of the trapped $CT_0$ state which lies on the $i^{th}$ molecule in the aggregate. The eigenstates of this Hamiltonian are found using a simple iterative procedure. We showed that while the maximally trapped $CT_0$ eigenstate becomes more localised and more strongly bound after relaxation occurs, the other higher lying states of the aggregate are essentially unaffected by relaxation, including those above the conduction edge which were shown to drive ultrafast charge separation. This modified bandstructure is illustrated in **Fig. S12.2b**.

These results motivate a simple description of the thermal separation of trapped pairs. We showed that higher lying states above the conduction edge can efficiently generate free charges on femtosecond timescales, while trapped pairs below this edge cannot undergo ultrafast separation. However if thermal fluctuations promote a trapped electron into the higher lying states before recombination occurs, then the charge pair will have an opportunity to separate. To estimate the time required to promote an electron into the higher lying states, we must specify the spectral function J(E), which describes the electron-phonon coupling. In our previous work we chose the simple Drude function $J(E) = \Delta E\gamma/(E^2 + \hbar^2\gamma^2)$, and we use the same spectral function here[30]. We show in a later section of the SI that the form of the spectral function does not significantly influence the dynamics, so long as the electron-phonon coupling is reasonably strong. The transition rate between two eigenstates of the aggregate can be estimated within Redfield theory[31,32],

$$R_{lm} = 2\pi J(|E_l - E_m|)n(E_l - E_m)\sum_i |C_l^i C_m^i|^2$$

The thermal function $n(E) = e^{E/kT}/(e^{E/kT} - 1)$ and $C_m^i$ describes the coefficient of the $m^{th}$ eigenstate on the $i^{th}$ lattice site, $|m\rangle = \sum_i C_m^i |i\rangle$.

We take a cubic aggregate of $4^3$ unit cells, a nearest neighbour coupling J = 25 meV [33], on site static disorder σ = 50 meV, dielectric constant $\varepsilon_r$ = 3.6 [13], an FCC lattice constant of 1.5 nm [34] and donor acceptor distance $r_{CT}$ = 1.5 nm [6]. The response time $\tau = 1/\gamma$ = 100 fs, and the reorganization energy Δ = 50 meV. In our previous study, we predicted that trapped electrons would be promoted into the higher lying states above the conduction edge within about 50 ps, however we will introduce a more robust treatment of the charge separation process in the following section.

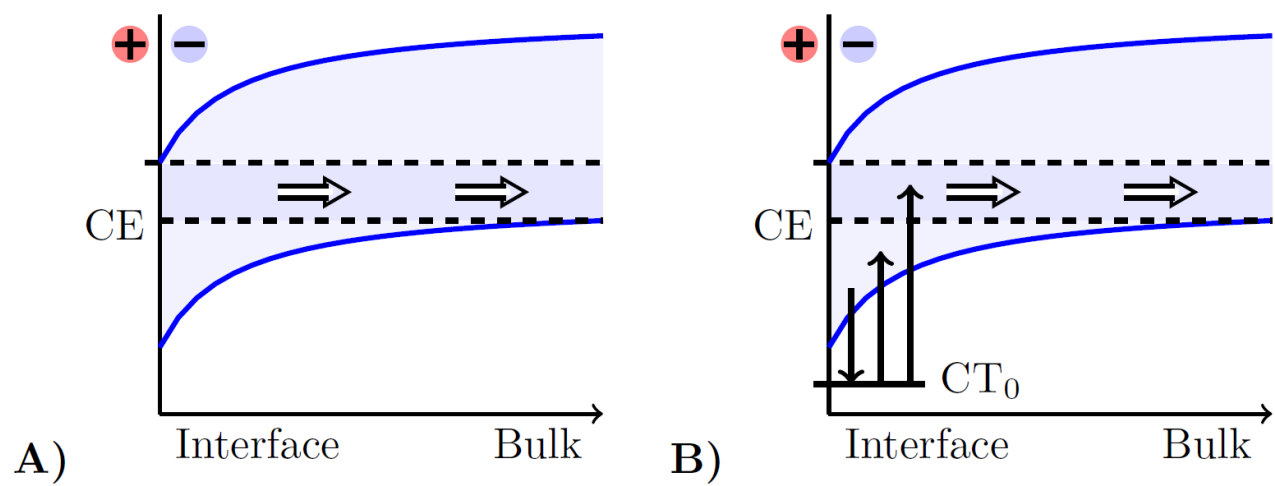

**Figure S12.2:** A) The eigenstates of the acceptor aggregate immediately after exciton dissociation, before vibronic relaxation can occur. A set of delocalised states exist above the conduction edge CE, which can drive ultrafast charge separation. B) On long timescales a polaronic $CT_0$ state forms, but the remaining higher lying states are largely unchanged. Thermal promotion to states above CE can drive charge separation on ps timescales.

### 12.3.2 A simple model of free charge generation

We now develop a simple model of free charge generation. We showed in previous work that eigenstates of the fullerene aggregate above the conduction edge can efficiently generate free charges on femtosecond timescales, while charges below this edge cannot. In the previous section we introduced the Redfield transition rates, which describe transitions within the aggregate and obey detailed balance.

$$R_{lm} = 2\pi J(|E_l - E_m|) n(E_l - E_m) \sum_i |C_l^i C_m^i|^2.$$

We supplement these rates by additionally assuming that any eigenstate above the conduction edge (energy $E_m > -12J$) generates free charges at a rate $R_{free} = 1/100$ fs$^{-1}$. Since we assume that equilibrium is strongly biased towards free charges and we are interested in the escape timescale for trapped pairs, we do not introduce a reverse process which regenerates electrons within the aggregate, and consequently the generation of free charges does not obey detailed balance. Additionally we do not include recombination, so all charges in the aggregate will eventually become free.

$R_{free}$ is chosen to be consistent with our earlier work on ultrafast charge separation[14], where the timescale for the generation of free charges from the higher lying states above the conduction edge was measured experimentally, and also observed within simulations. In **Fig. S12.3** we exhibit the population of free charges under two initial conditions. "Cold" injection places the electron in the maximally trapped CT$_0$ state of H$_{eff}$ at time t = 0. This mimics the dissociation of low-energy triplet excitons. "Hot" Injection places the electron with equal probability in all eigenstates of the aggregate, more similar to the dissociation of high-energy singlet excitons. "Hot" injection leads to substantial ultrafast charge separation on femtosecond timescales, followed by a continued slow rise in the free charge population. "Cold" injection leads to free charge formation on picosecond timescales, in excellent agreement with our experimental results.

The initial growth of free charges depends strongly on the initial conditions. However at long times, the ratios of the populations in the aggregate eigenstates are constant, while the magnitude of the population in all the aggregate eigenstates slowly decays. In this limit, the population of free charges obeys $P_{free}(t) = 1 - e^{-Rt}$, where $R = \alpha/100$ fs$^{-1}$ and $\alpha$ is the population fraction above the conduction edge. We can find this steady state as follows. We form the matrix of transition rates $R_{lm}$, excluding the free charge state. We account for the free charge state, while ensuring that the population inside the aggregate remains constant, by increasing the transition rate from states above the conduction edge CE down to the ground CT$_0$ state by $1/100$ fs$^{-1}$. We may then solve for the steady state population vector p, which satisfies $\sum_m R_{lm} p_m = 0$, from which we can trivially obtain $\alpha$. Any individual realisation of the system will be described by mono-exponential growth, but the rate R is disorder dependent, and the final population curve $\langle P_{free}(t) \rangle$ is an average across a range of simple exponential rates.

We also include the free charge population under this steady state method in **Figure S12.3**. We see that the method provides an extremely good approximation to the free charge population generated from the ground $CT_0$ state, especially on timescales longer than 1ps. This steady state approximation is significantly more numerically efficient. Note that in order to generate free charge on a timescale $\tau_{free} = 1/R < 100$ ps, we only require that the fraction of electron density above the conduction edge, $\alpha > 0.1\%$. This rationalizes how an efficient high lying escape channel can enable charge separation before recombination, even in the presence of a deep trap.

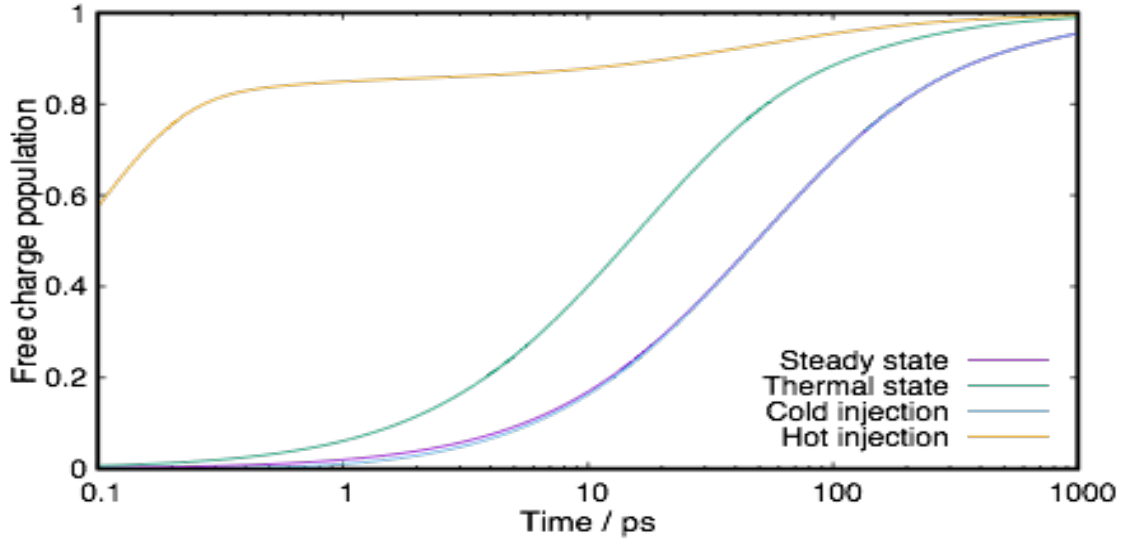

**Figure S12.3:** The hot and cold injection curves describe numerical integration of the rate equations under the two initial conditions described in section 12.2. $R_{free} = 1/100$ fs$^{-1}$. The steady state curve follows the approximate method described in section 12.2, while the thermal state curve follows the method of section 12.3. The steady state method provides an excellent approximation to the cold injection curve on timescales longer than a few ps.

### 12.3.3 The thermal approximation

The dynamical equations described in the previous section appear to depend crucially on the choice of spectral function J(E). To illustrate that this is not always the case, we introduce a second method to estimate the growth in the free charge population: we assume that the eigenstates of the aggregate reach internal thermal equilibrium at temperature T. The population of any eigenstate in the aggregate, $P_m = e^{-E_m/kT}/Z$, where $Z = \sum_n e^{-E_n/kT}$. Once again, $P_{free}(t) = 1 - e^{-Rt}$, where $R = \alpha/100$ fs$^{-1}$ and $\alpha$ is the population fraction above the conduction edge. This method enables us to estimate the timescale to generate free charges solely from the energy spectrum, without specifying any transition rates or spectral function.

We included the thermal state curve in **Fig. S12.3**, where it overestimated the rate at which free charges form from the CT$_0$. This can be simply understood. All transitions inside the aggregate obey detailed balance, but the generation of free charges from states above the conduction edge does not. The generation of free charges slightly depletes the populations above the conduction edge below their equilibrium values, decreasing the rate at which free charge is generated. The thermal state approximation thus provides an upper bound, predicting the maximum rate at which bound CT states can generate free charges within a given eigenstate spectrum.

To emphasize this, in **Fig. S12.4** we plot dynamics for a range of free charge generation rates $R_{free}$. When $R_{free}$ is small (Fig S12.4b), it only provides a small perturbation and the eigenstate populations are close to thermal. In this limit the thermal state approximation is a good approximation to cold injection. When $R_{free}$ is large (Fig. S12.4a), it provides a significant perturbation, and the eigenstate populations are far from thermal. The accuracy of the thermal approximation is controlled by the relative scale of the transition rates inside the aggregate to $R_{free}$. We could also improve the accuracy of the thermal approximation by increasing the reorganisation energy, which sets the overall strength of the electron-phonon coupling within the aggregate. However this changes $H_{eff}$, leading to a modified bandstructure.

By contrast, hot injection can only lead to ultrafast free charge generation if $R_{free}$ is larger than the rate at which the aggregate internally thermalises. The smaller $R_{free}$, the smaller the difference between the hot and cold injection curves.

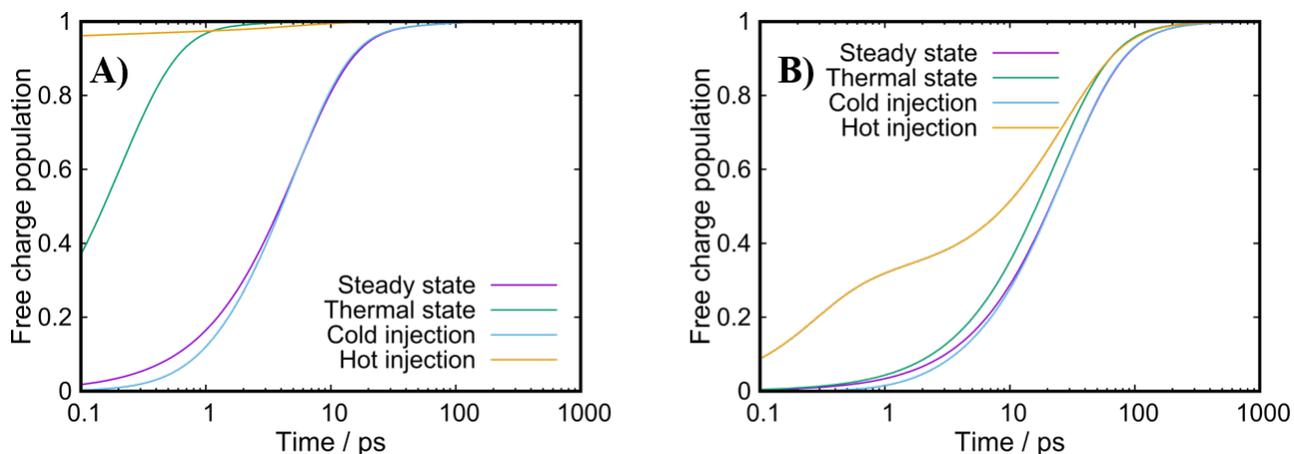

**Figure S12.4**: A) Dynamics in an identical aggregate, with modified free charge generation rate $R_{free}$ = 1/10 fs$^{-1}$. The steady state approximation remains excellent. The ultrafast free charge component for hot injection is substantially enhanced, and the gap between the thermal approximation and the true dynamics for cold injection has increased. B) $R_{free}$ = 1/1000 fs$^{-1}$. The ultrafast component under hot injection is substantially reduced. Free charge generation only weakly perturbs the eigenstate populations, and the thermal curve, steady state curve, and cold injection curve are all relatively similar. We observe that even when the timescale for hot electrons to escape the aggregate is slow, the majority of charges have still escaped within 100 ps.

### 12.3.4 Predicting the electro-absorption signature

We cannot directly measure the free charge population, instead we measure the EA signature; which tracks the averaged squared electric field experienced by donor molecules inside the blend [14,35]. Thus, $EA(t) \sim \propto \int_{V_D} |E(r,t)|^2 \, dr^3$, where α is a constant and $V_D$ denotes the donor regions of the blend. We assume that the integrated squared electric field across donor and acceptor regions are equal, and approximate that the dielectric constant is uniform throughout the device. Under these approximations, $\varepsilon_0 \epsilon_r EA(t) \sim \alpha \int_V |E(r,t)|^2/2 \, dr^3$, where V denotes the entire blend volume.

$\varepsilon_0\varepsilon_r|E(\underline{r},t)|^2/2$ is the electrostatic energy density stored at the point $\underline{r}$ in the field at time t. Consequently EA(t) = αV(t)/$\varepsilon_0\varepsilon_r$, where V(t) denotes the electrostatic potential of the charge pair. The EA contribution of two neighbouring charges is absorbed into existing charge signatures, so the electrostatic potential is normalised to zero when the electron and hole lie $r_{CT}$ = 1.5 nm apart. V(t) tracks the work done against the Coulomb potential in separating the electron and hole beyond this nearest neighbour separation.

The electrostatic potential of the m$^{th}$ eigenstate, $V_m = W - q^2 \sum_i (|C_m^i|^2 / 4\pi\varepsilon_0\varepsilon_r r_i)$, where W = $q^2/4\pi\varepsilon_0\varepsilon_r r_{CT}$, and the electrostatic potential of a free charge $V_{free}$ = W. Thus the electrostatic potential of a single disorder realization, $V(t) = p_{free}(t) W + (1 - p_{free}(t)) G$, where $G = \sum_m p_m V_m$. $p_m$ labels the normalised steady state population vector discussed in section 12.3.2, and G < W. For a single disorder realisation,

$$V(t) = G + (W - G)(1 - e^{-Rt})$$

R is calculated using the steady state approximation. We take $R_{free}$ = 1/100 fs$^{-1}$, consistent with experimental observations of ultrafast charge separation. In **Fig. S12.4a** we plot the electrostatic potential V(t) following formation of a CT state in red. Since the CT$_0$ is delocalised, we observe substantial potential even at t = 0. As free charges form, the potential saturates at the Coulomb well depth $W \sim 270$ meV.

Experimentally we saw bi-exponential CT state formation, $P_{CT}(t) = 1 - Ae^{-R_1 t} - (1-A)e^{-R_2 t}$. The rate of CT state formation, $R_{CT}(t) = dP_{CT}/dt = AR_1 e^{-R_1 t} + (1-A)R_2 e^{-R_2 t}$. The predicted EA signature $EA(t) = \int_0^t R_{CT}(t')V(t-t')dt'$. Evaluating the convolution,

$$EA(t) = A\left\{W(1-e^{-R_1 t}) - \frac{R_1(W-G)(1-e^{-(R+R_1)t})}{R+R_1}\right\}$$
$$+(1-A)\left\{W(1-e^{-R_2 t}) - \frac{R_2(W-G)(1-e^{-(R+R_2)t})}{R+R_2}\right\}$$

This expression must be averaged over the disorder distribution of the charge generation rate R. We plot this predicted EA signature in **Fig. S12.4a** in green. For simplicity we continue to neglect recombination. The EA saturates within around 50 ps, consistent with our experimental results. To test the resilience of our model, in **Fig. S12.4b** we perform the same set of simulations for a second system where the separation distance between donor and acceptor is reduced to $r_{CT} = 1.0$nm, such that the Coulomb well depth $W \sim 400$ meV. We continue to observe a substantial EA on 10-1000 ps timescales.

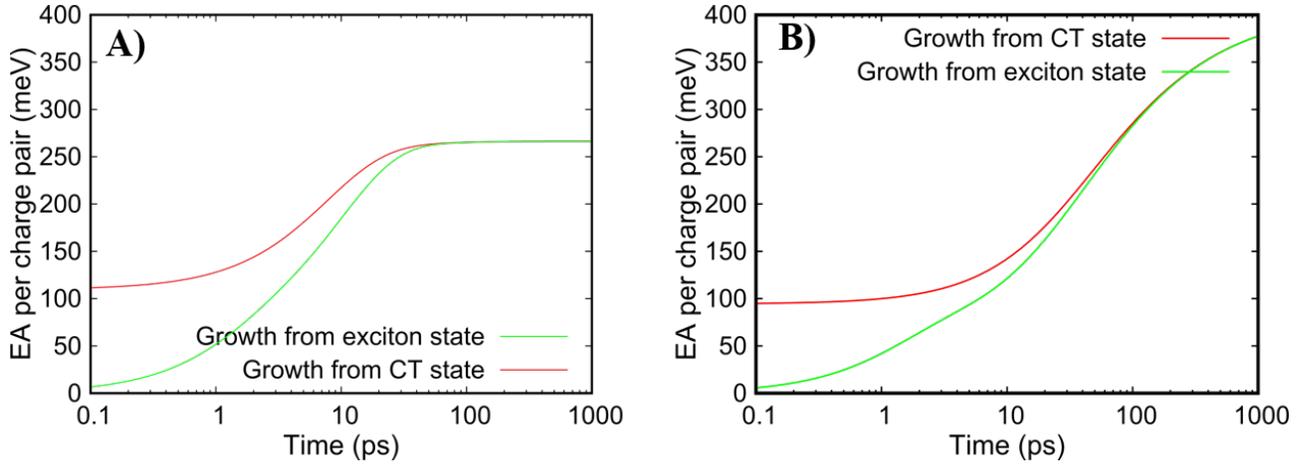

**Figure S12.5. A:** The red curve tracks the growth in the stored electrostatic energy, following the formation of a CT state. This is the quantity measured by the EA signature. The green curve convolves the red curve with the rate at which triplet excitons dissociate into CT states, and is thus directly comparable to our experimental data. As observed in experiment, we assume that CT states form on two timescales ($\tau_{fast}$ = 1.2 ps and $\tau_{slow}$ = 8.9 ps) in a 2.4:1 ratio. Since we neglect recombination, this curve is only reliable for the first 100 ps or so. The green curve is also shown in the main text. **B:** Here we plot a similar set of results for an identical model,

except the donor acceptor separation $r_{CT}$ = 1.0 nm, increasing the depth of the Coulomb well to ~ 400 meV.

## 12.4  CT state dissociation within Marcus theory

### 12.4.1  Dynamics within Marcus theory

Within Marcus theory[36], charges are localised on lattice sites. A charge remains on a given lattice site, until at some later time it spontaneously hops onto a neighbouring site. The rate at which a given charge hops from state i to state j is given by

$$k_{ij} = \frac{2\pi J^2}{\hbar\sqrt{4\pi\Delta k_B T}} e^{-(E_{ij}+\Delta)^2/4\lambda k_B T}$$

$E_{ij}$ denotes the difference in free energy between state i and state j. This energy may depend on interactions with the other particles in the system. J denotes the nearest neighbour coupling strength between the two states, and Δ represents the reorganization energy of a charge on a given lattice site. We take a simple model where both the donor and acceptor phases have a primitive cubic unit cell with a lattice constant of 1 nm. Furthermore, we take our device to be composed of a thin slab 8 unit cells thick perpendicular to the interface, 4 of which are donor and 4 of which are acceptor. The width of the slab in the two directions parallel to the interface is 1000 unit cells. The separation between donor and acceptor sites at the interface, $r_{CT}$ = 1.5 nm.

The coupling constants in fullerene are comparatively uniform in three dimensions[33], while pentacene forms crystals of stacked planes[37]. The couplings in the plane are comparatively strong, while the couplings perpendicular to the planes are weak. We model this by taking the nearest-neighbour coupling constants in the acceptor phase to be 50 meV in all directions, while the coupling constants in the donor phase are 50 meV in the two directions parallel to the interface, and 5 meV in the direction perpendicular to the interface[37]. We take the reorganization energy of both donor and acceptor to be 50 meV.

We initialize the system in a CT state, whereby the electron and hole lie on neighbouring sites across the interface in the centre of our slab. We assume that the electron and hole can never reform an exciton, and only allow the hole to hop within the donor phase and the electron to hop within the acceptor. The energy of a state is given by the static disorder experienced by the electron and hole, as well as the Coulomb potential experienced between the two charges. If the electron lies on site a and the hole lies on site b, the energy

$$E_{ab} = \sigma_a + \sigma_b - e^2/4\pi\varepsilon_0\varepsilon_r r_{ab}$$

Where $\sigma_a$ and $\sigma_b$ represent the static disorder on the $a^{th}$ electronic site and $b^{th}$ hole site. For simplicity we set the standard deviation $\sigma = 50$ meV for both electrons and holes. e denotes the electron charge, and $r_{ab}$ denotes the distance between site a and site b. We take the dielectric constant $\varepsilon_r = 3.6$. Since we fixed the minimum distance between electron and hole to be 1.5 nm, corresponding to an Coulomb well of ~270 meV.

The dynamics within Marcus theory can be simulated within the Kinetic Monte Carlo framework[38]. Assume that the electron and hole lie in a specific state i =¹ at time t. We can evaluate the rates $k_{ij}$ for all possible nearest neighbor hops from state i to states j. The total rate $k = \sum_j k_{ij}$. We generate a random number u, uniformly distributed between 0 and 1, and we update the time $t \to t - \ln u /k$, and we simultaneously update i to a new state $j_{new}$ selected randomly from the states j, where the probability of choosing a given state j is proportional to $k_{ij}$. Repeating this procedure, we generate one possible dynamical trajectory for the charge pair. We must run the simulation many times to average across these trajectories.

### 12.4.2 Predicting the Electro-absorption signature

At each time-step we can calculate the EA of the state, $EA_{ab} = (e^2/4\pi\varepsilon_0\varepsilon_r)(1/r_{CT} - 1/r_{ab})$. We fixed the charge separation of the CT state, $r_{CT} = 1.5$ nm above. The dynamics of the EA provides an estimate of the rate at which CT states dissociate into free charges. We plot the EA in **Fig. S12.6a**, where we have averaged across x individual trajectories. We see that a substantial EA is generated within the first 100 ps, consistent with our experimental observations. Although the model is extremely simplified and does not account for the real crystal structures of pentacene and $C_{60}$, all the parameters chosen above are within the plausible range, demonstrating that CT state dissociation within 10-100 ps is consistent with a Kinetic Monte Carlo treatment, as well as being consistent with the delocalised state model described in the previous section.

In order to compare these simulations to the observed experimental EA signature, we must also take account of formation of CT states from excitons. Experimentally we observed that CT state formation followed a bi-exponential curve, $P_{CT}(t) = 1 - Ae^{-R_1 t} - (1-A)e^{-R_2 t}$. We can account for this within our simulations by beginning the simulations in the exciton state. We draw two random numbers v and q, both between 0 and 1. If v < A, then the initial CT state is formed at time $t = -\ln(q)/R_1$. If v > A, then the initial CT state is formed at $t = -\ln(q)/R_2$. We also exhibit this curve for the overall EA signature in **Fig. S12.6a**, and it was also shown in the main text. As for the quantum model, this curve is only reliable within the first 100 ps, since we do not account for recombination.

In **Fig. S12.6b**, we perform the same KMC simulation, for the case $r_{CT}$ = 1 nm; which implies that the binding energy is ~ 400 meV. In this case the observed EA is much smaller, as the electron is much more strongly bound to the hole. This contrasts with the delocalised dynamics model, where a substantial EA was still seen for deeper binding energies. Charge separation within KMC is less resilient than that predicted by delocalised dynamics, despite the fact that the classical model allows both charges to move, while the delocalised dynamics only considered electron motion. However we cannot exclude either model, since both sets of parameters lie within a plausible range.

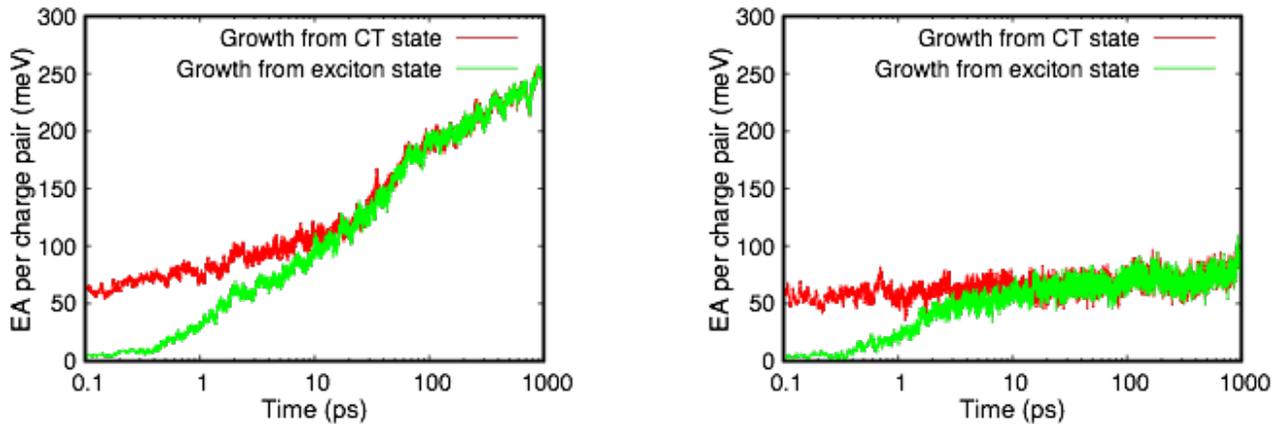

**Figure S12.6:** The EA under Marcus dynamics. A) A small EA has already emerged 0.1 ps after exciton dissociation, and it continues to rise over a nanosecond. The green curve accounts for the dissociation of triplets into CT states on two timescales (τ_fast = 1.2 ps and τ_slow = 8.9 ps) in a 2.4:1 ratio. B) The evolution of the EA when $r_{CT}$ = 1 nm, implying a binding energy of ~ 400 meV. The slow long-ranged separation of electron and hole on 10-1000 ps timescales is suppressed.